\documentclass[notitlepage,nofootinbib,
twocolumn,
longbibliography,superscriptaddress,amsmath,amssymb]{revtex4-1}

\usepackage{graphicx}  % needed for figures
\usepackage{dcolumn}   % needed for some tables
\usepackage{bm}        % for math
\usepackage{verbatim}   % for math
\usepackage{bbold}
\usepackage{color}
\usepackage{xcolor}
\usepackage{amsthm}
\usepackage{amsmath}
\usepackage{ulem}
\usepackage{eufrak}

\usepackage{amsmath}
\usepackage{amsthm, amssymb}
\usepackage{slashed} 
\usepackage{graphicx}
\usepackage{xcolor}
\usepackage{bm}
\usepackage[colorlinks=true,citecolor=blue,linkcolor=magenta]{hyperref}
\usepackage{dsfont}
\usepackage{changepage}
\usepackage{array}

\def\be{\begin{equation}}
\def\ee{\end{equation}}
\def\ba{\begin{align}}
\def\ea{\end{align}}
\def\>{\rangle}
\def\<{\langle}

\def\bea{\begin{eqnarray}}
\def\eea{\end{eqnarray}}
\def \T*{\text{T}^*}
\def \Tc{\text{T}_\text{c}}
\def \K3C60{\text{K}_3\text{C}_{60}}

\def\>{\rangle}
\def\<{\langle}

\theoremstyle{definition}

\theoremstyle{remark}

\begin{document}

\title{Photo-induced superconducting-like response in strongly correlated systems}

\author{Zhehao Dai}
\affiliation{
Department of Physics, University of California, Berkeley, CA 94720, USA}
\affiliation{
Department of Physics, Massachusetts Institute of Technology, Cambridge, MA 02139, USA
}
\author{Patrick A. Lee}
\affiliation{
Department of Physics, Massachusetts Institute of Technology, Cambridge, MA 02139, USA
}

\date{\today}

\begin{abstract}
We propose a novel mechanism for the photo-induced superconducting-like response recently reported in cuprates and other strongly correlated materials. This mechanism relies on quantum-fluctuating bosons consisting of electron pairs. With periodic drive, the electron pairs and vacancies of pairs form a coherent non-equilibrium condensate, different from conventional superconductors, yet showing superconducting-like response in some regime even with dissipation. Unlike the case of driven fermionic bands which results in the familiar Floquet bands with hybridization gaps, for driven bosons the ``gap" opens up in the momentum direction, resulting in a resonant region in momentum space where the eigenvalues are complex. We give a simple physical argument why this picture leads to a``perfect conductor" which exhibits superconducting-like frequency-dependent conductivity but no Meissner response.  While our model is quite general, in the case of cuprates, quantum-fluctuating pair density wave in the pseudogap region may serve as the origin of the quantum-fluctuating electron pairs.
\end{abstract}

\maketitle

\section{Introduction}

In the last ten years, photo-induced superconducting-like response have been reported in various strongly correlated materials, including multiple species of cuprate high-temperate superconductors, $\K3C60$, and an organic superconductor~\cite{fausti2011light,PhysRevB.89.184516,hu2014optically,mitrano2016possible,cavalleri2018photo,PhysRevX.10.031028}. In these pump-probe experiments, a pump pulse excites the sample; the transient ac conductivity is then inferred by measuring the reflection coefficients of a probe light. Very surprisingly, samples originally in the \textit{normal} phase behave like superconductors for picoseconds after the pump. Most recently, the lifetime of the transient superconductivity is extended to nanoseconds~\cite{budden2020evidence}, which allows a direct measurement of the two-terminal resistance.

The microscopic mechanism of the photo-induced superconducting-like response is still unclear, and it may depends on details of the material. In $\text{La}_{1.8-x}\text{Eu}_{0.2}\text{Sr}_{x}\text{CuO}_4$ (LESCO) and $\text{La}_{2-x}\text{Ba}_x\text{CuO}_4$ (LBCO), the response is found near 1/8 doping, where superconducting transition temperature $\Tc$ in equilibrium is suppressed to almost zero by `stripe orders'~\cite{fausti2011light,PhysRevLett.112.157002,cavalleri2018photo}. X-ray diffraction experiments reported the melting of the stripes after the pump, which is argued to be connected to the transient superconducting response below a temperature scale comparable to the equilibrium $\Tc$ of nearby doping. However, this mechanism cannot apply to $\text{Y}\text{B}_2\text{Cu}_3\text{O}_y$ (YBCO), where the temperature T$'$ below which the photo-induced superconducting-like response is found, coincides with the pseudogap temperature $\T*$ for multiple underdoped samples. At 7\% hole doping, the transient superconductivity exists up to at least 300K, which is 8 times larger than the equilibrium superconducting transition temperature $\Tc= 35$K~\cite{cavalleri2018photo}. A more detailed study~\cite{PhysRevX.10.011053} shows that the superconducting-like response is induced only when the pump frequency is close to the oscillation frequencies of the apical oxygen, suggesting the change in the electronic structure is driven by the resonantly excited optical phonon. In $\K3C60$, the photo-induced superconducting-like response also exists below a temperature T$'\gg\Tc$, but the response is seen over a broad frequency range of the pump light.~\cite{nava2018cooling,cavalleri2018photo}

The experiments have stimulated many theoretical efforts. Ref.~\cite{PhysRevB.91.094308,PhysRevB.93.144506} discussed the possibility that a transient static lattice distortion, resulting from nonlinear phonon couplings, favors higher superconducting transition temperature. Ref.~\cite{PhysRevLett.114.137001,nava2018cooling} discussed potential cooling effects of the pump. Ref.~\cite{komnik2016bcs,kennes2017transient,PhysRevB.94.214504,PhysRevB.96.014512,PhysRevB.96.045125} discussed the pump-enhanced electron-phonon interaction and its impact on superconductivity. Ref.~\cite{vonhoegen2020parametrically,kleiner2020spacetime,homann2020higgs,PhysRevLett.117.227001} discussed the enhancement of superconducting response through parametric amplification of Josephson plasma modes and/or the Higgs mode. Ref.~\cite{lemonik2019transport} discussed superconducting fluctuations and its experimental signatures induced by the pump in a normal state. Going beyond superconductivity, hence perfect reflection of the probe light, Ref.~\cite{PhysRevX.11.011055,PhysRevB.102.174505} discussed the enhancement of the probe light, Ref.~\cite{PhysRevB.98.220510} discussed transient negative conductivity.

In this paper, we propose a new mechanism for photo-induced superconducting-like response. We focus on the limit $\Tc \ll \text{T} \ll \text{T}',\Omega$, where T is the temperature of the sample and $\Omega$ is the frequency of the pump. Since $\text{T} \gg\text{T}_c$, the equilibrium superconducting order and the thermal fluctuations of the Cooper pairs do not play a role; on the other hand $\text{T} \ll \text{T}',\Omega$, so that the system is at low temperature compared to the energy scale relevant to the physics of the `driven superconductivity'. 

In this limit, we assume that the response comes from electron pairs which do \textit{not} condense at equilibrium due to quantum fluctuations. We show that a periodic drive produces an intrinsically non-equilibrium state of the electron pairs different from conventional superconductors; nonetheless, the state shows coherent superconducting-like electromagnetic response.

For YBCO, the assumed quantum-fluctuating electron pairs in the pseudogap region may come from fluctuating pair density waves (PDW), which is studied in details in a recent  work of the authors'~\cite{PhysRevB.101.064502}. In this proposal, electrons near the antinodes are gapped by pairing at nonzero momenta. However, at low temperature, the electron pairs form a bosonic Mott insulator instead of a superfluid in the unit cell enlarged by the charge density wave. Thus, apart from quasi-electron excitations near the node, we also have gapped excitations of electron pairs and vacancies of electron pairs near the PDW momentum, which are described by the following Lagrangian

\bea
\mathcal{L} = \frac{1}{2}|\partial_t \Psi|^2 - \frac{v^2}{2}|\nabla \Psi|^2 - \frac{1}{2}\Delta^2|\Psi|^2 - \frac{U}{8}|\Psi|^4,
\label{Eq: boson Lagrangian without drive}
\eea
where $\Psi$ is a complex boson field; $v$, $\Delta$, and $U$ are the velocity, gap, and interaction strength of the pair.  
%These non-condensing pairs (coexisting with unpaired electrons) are recently proposed from the perspective of fluctuating pair density waves, to explain the mysterious pseudogap phenomenology in cuprates~\cite{PhysRevB.101.064502}. 

For the propose of this work, the existence of bosonic particle and hole excitations is important, but the exact ground state is not. We shall proceed phenomenologically using the Lagrangian above. As a qualitative description of the electron pairs, Eq.~\ref{Eq: boson Lagrangian without drive} may apply to both 2D and 3D materials; with small changes, it also describes phenomena specific to bilayer cuprates, which we discuss in a separate work~\cite{bilayer}. We emphasize that in contrast to previous works that discuss how an existing superconductor can be enhanced by a periodic drive, we show that even an insulator can have superconducting-like response under periodic drive.

To describe the non-equilibrium physics after the pump, we add a periodic modulation to the boson gap (see Sec.~\ref{Sec: model, quantum state} for details), which may come from the resonantly excited optical phonon in YBCO, or directly from the pump in the recent experiment on $\K3C60$, where laser pulses with much longer duration are used~\cite{budden2020evidence}. We also discuss an alternative model consisting of non-relativistic bosons (Eq.~\ref{Eq:nonrelativisticLagrangian}), which captures the essential non-equilibrium physics of the relativistic model, but is much easier to handle analytically. 

In both models, the periodic drive resonantly excite particle-hole pairs of the boson, resulting in a time-dependent condensate of the electron pairs and vacancies of electron pairs. The density of the excitations grows exponentially at early time, and finally saturates due to the nonlinear interaction if the periodic drive continues indefinitely. In real experimental settings, the evolution of the condensate depends on the total energy of the pump, the strength of the nonlinear interaction and the dissipation rate of the boson. We discuss several scenarios in Sec.~\ref{Sec: late time behavior}. We show that the ac conductivity of the condensate is like that of a superconductor, $\sigma(\omega)\propto i/\omega$, even with dissipation. This $1/\omega$ behavior is cut off only when $\omega$ approaches the inverse of the duration of the photo-induced state, namely the smallest frequency one can resolve in the transient phenomena. We give analytic results for ac conductivity at early time (Sec.~\ref{Sec: linear response} and Sec.~\ref{Sec: dissipation}) and numerical results for ac conductivity beyond early time. 

On the other hand, we find that there is \textit{no} Meissner effect for the response to a static magnetic field at early time. So the phenomenology is quite different from that of a superconductor. It is closer to a free fermion gas without any source of scattering or dissipation, but of course we find this behavior even in the presence of dissipation. We shall refer to this phenomenon as a ``perfect conductor".  It is only at late time, if the excited electron pairs have enough time to relax to the zero momentum before the energy of the pump is completely dissipated, that we expect a Meissner effect. In that scenario, the driven system may arrive at a novel non-equilibrium steady state at late time, which is not likely the case for experiments performed on YBCO but which may be the case in the recent experiment on $\K3C60$~\cite{budden2020evidence}. We shall discuss this steady state and its electromagnetic response in a separate work~\cite{steadyState}.

%This work reveals an intrinsically non-equilibrium mechanism for superconductivity, which clearly deserves further investigations.

\section{The driven boson model and the evolution of the quantum state}
\label{Sec: model, quantum state}

\begin{figure}[htb]
\begin{center}
\includegraphics[width=0.25\textwidth]{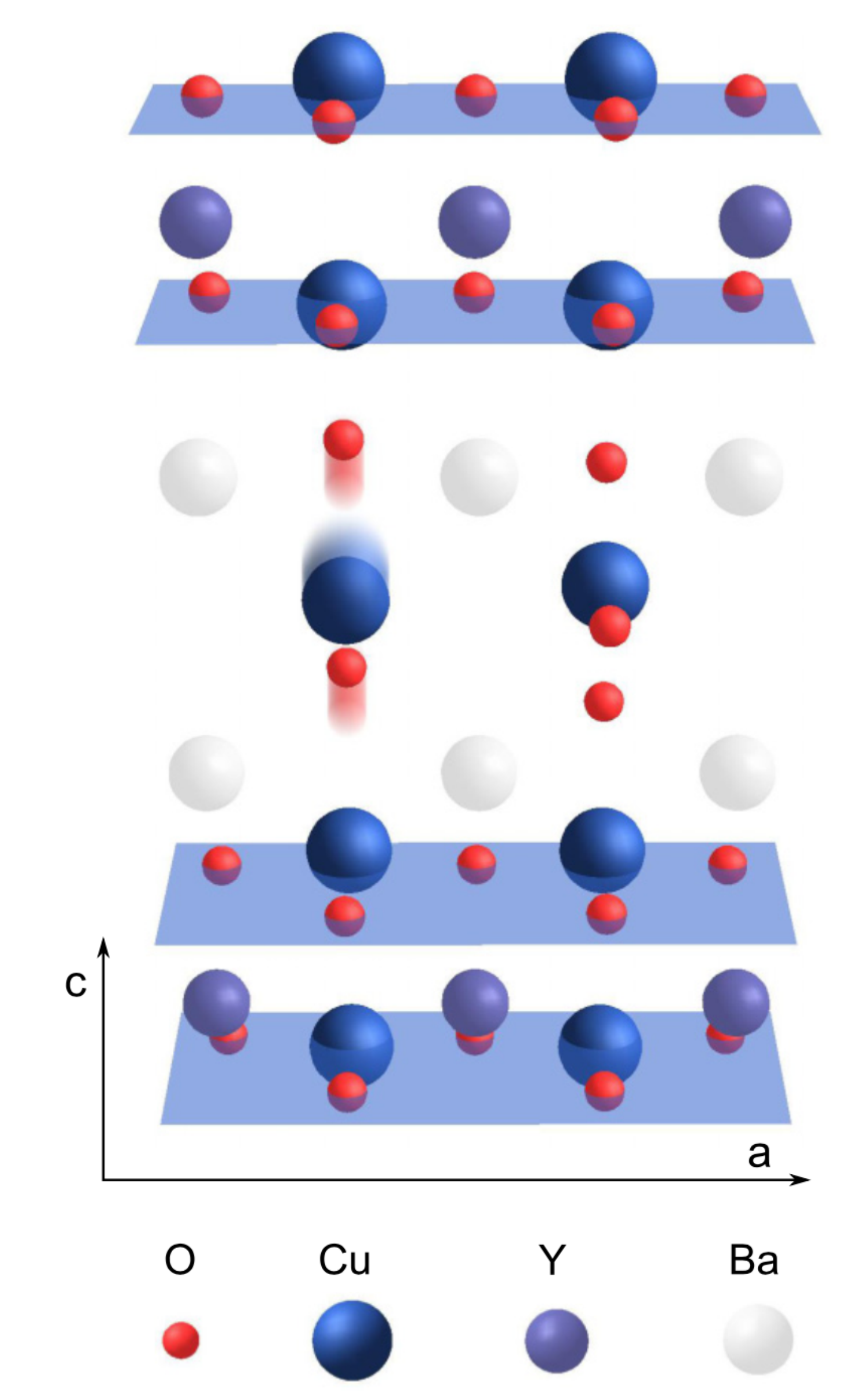}
\caption{Reproduction of Fig. 1(c) in Ref.~\cite{PhysRevB.91.094308}. The THz pump excites the c-axis oscillation of the apical oxygens.}
\label{Fig:YBCO}
\end{center}
\end{figure}

In this section, we discuss the periodically driven boson model and the initial time evolution of the quantum state. The relativistic boson model we consider is 

\begin{align}
\mathcal{L}_\text{rel} =& \frac{1}{2}|\partial_t \Psi|^2 - \frac{v^2}{2}|\nabla \Psi|^2 - \frac{1}{2}\Delta^2|\Psi|^2 - \frac{U}{8}|\Psi|^4\nonumber\\
& - \lambda\cos(\Omega t)|\Psi|^2.
\label{Eq: relativistic Lagrangian with drive}
\end{align}

The driving term we choose is not unique, but represents one of the simplest possibilities. In real experiments, the dominant driving term depends on materials. YBCO, for example, is composed of $\text{CuO}_2$ bilayers (adjacent blue planes in Fig.~\ref{Fig:YBCO}) and atoms in between. We assume that each layer contains electron pairs described by Eq.~\ref{Eq: boson Lagrangian without drive}~\cite{PhysRevB.101.064502}, where the parameters depends on the positions of the atoms between the bilayers. To the leading order, the resonantly excited oscillation of the oxygen illustrated in Fig.~\ref{Fig:YBCO} should induce opposite changes of the parameters in the upper and lower $\text{CuO}_2$ layers in each bilayer, for example, a modulation of the boson gap described in Eq.~\ref{Eq: relativistic Lagrangian with drive}. In that case, the coefficient $\lambda$ is proportional to the oscillation amplitude of the oxygen. For simplicity, we focus on a single layer in the current work, which already captures the essence of the non-equilibrium physics. In addition to the single-layer physics, the weak c-axis coupling in YBCO gives new signatures in the c-axis response, which we discuss in a separate work~\cite{bilayer}. For a single layer, in addition to the periodic modulation of the gap, one may also consider a periodic coupling to the charge density. However, to the leading order, this driving term leaves the ground state invariant since the total charge commutes with the Hamiltonian.

We may also apply similar models to isotropic 3D materials. In that case, if the material has an inversion symmetry or mirror reflection symmetry, the driving term in Eq.~\ref{Eq: relativistic Lagrangian with drive} is forbidden in the first order of the pump electric field. We can replace the driving term by a periodic coupling to the current density. We shall see later in this section that the early time behavior is qualitatively similar to that of Eq.~\ref{Eq: relativistic Lagrangian with drive}.

Now we analyze the evolution of the bosons at early time, starting from its ground state. We define the early time as the period when the phonon amplitude have not significantly decayed, so that $\lambda$ is approximately constant, and when the excited boson amplitude is relatively small, $|\Psi|^2 < \Delta^2 /U, \lambda/U$, such that the boson interaction is not important. In this limit, we find an analytic solution of the time evolution.

\begin{figure*}[htb]
\begin{center}
\includegraphics[width=0.6\textwidth]{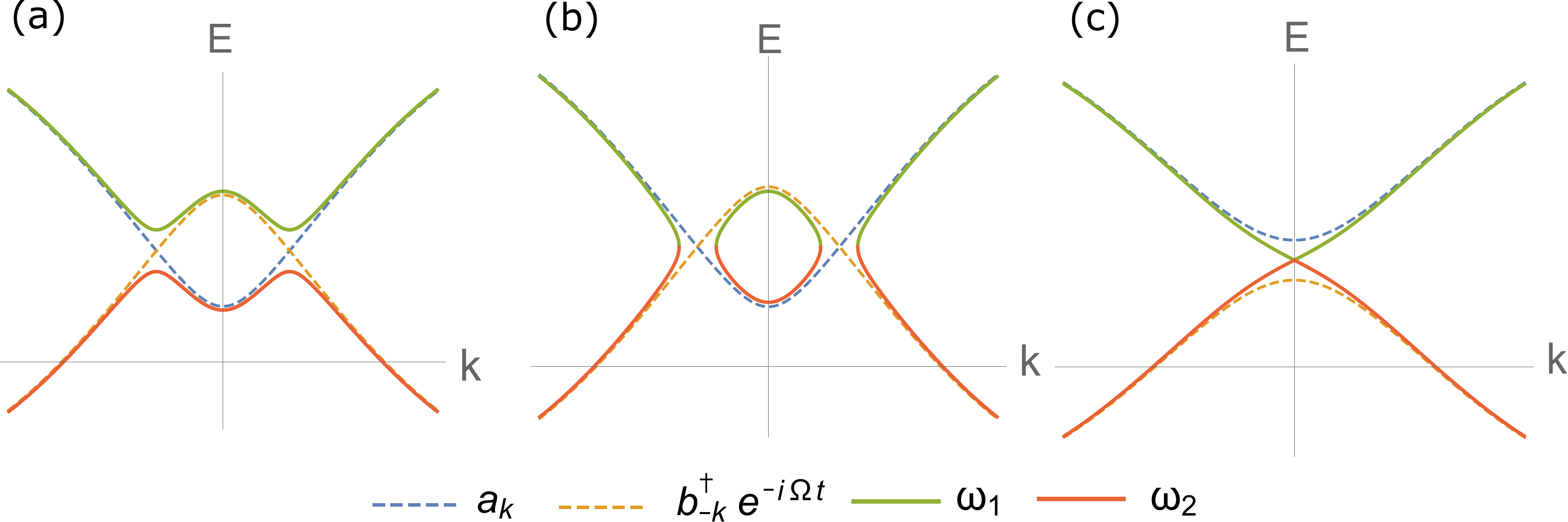}
\caption{(a) The hybridization of two fermion bands opens an energy gap.(b) Hybridization of two boson bands. The blue dashed line is the dispersion of $a_k$ in the free boson model, the yellow dashed line is the dispersion of $b_{-k}^\dagger$ shifted by $\Omega$, and the solid lines are the eigen-frequencies of the driven system. Contrary to the fermion case, there is a momentum range without real eigen-frequencies. (c) Hybridization of the boson bands at the critical gap, when $\Delta = \Omega/2 +\lambda/\Omega$. The effective dispersion after the hybridization is linear near $\Omega/2$.}
\label{Fig:drivenDispersion}
\end{center}
\end{figure*}

To provide some intuition, we first consider the classical equation of motion.

\be
\partial_t^2\Psi_k = (E_k^2 + 2\lambda\cos(\Omega t))\Psi_k,
\ee
where $\Psi_k$ is the spatial Fourier transform of $\Psi$ at time t,
\be
E_k \equiv \sqrt{\Delta^2 + v^2k^2}
\ee
Without the periodic drive, $\Psi_k$ oscillates at its natural frequency, $\pm E_k$. When $\Omega\simeq 2E_k$, the periodic drive resonantly mixes these two modes, resulting in an exponentially growing amplitude of the oscillation, known as parametric resonance. In the quantum model, the exponential growth starts from the zero-point fluctuation of $\Psi_k$. In fact, simple models of parametric resonances is well-studied in the early days of quantum optics~\cite{PhysRev.160.1076}.

Let us perform standard canonical quantization to the time-dependent Lagrangian,

\be
[\Pi(x),\Psi(x')] = -i\delta(x-x'),
\ee
where 
\be
\Pi \equiv \frac{\delta \mathcal{L}_\text{rel}}{\delta\partial_t\Psi} =  \frac{1}{2}\partial_t\Psi^*
\ee
Define, 
\begin{align}
\label{Eq:canonicalQuantization}
\Psi_k &\equiv \frac{1}{\sqrt{E_k}}(a_k + b_{-k}^{\dagger})\\
\partial_t\Psi_k &\equiv \sqrt{E_k}(-i a_k + ib_{-k}^{\dagger}) &
\end{align}
It is easy to check $a_k$ and $b_k$ obey the commutation relations of independent boson annihilation operators. We identify $a_k$ as the annihilation operator of electron pairs, and $b_k$ as the annihilation operator of vacancies of electron pairs relative to a background. And the Hamiltonian is given by
\be
H_t = \sum_k E_k(a_k^\dagger a_k + b_{k}^\dagger b_k) + \frac{\lambda}{E_k}(a_k^\dagger + b_{-k})(a_k+b_{-k}^\dagger)\cos(\Omega t)
\label{Eq:relativisticHamiltonian}
\ee

In the Heisenberg picture, an operator $O$ evolves as
\be
dO(t)/dt = i[H_t(t), O(t)],
\label{Eq:Heisenberggeneral}
\ee
where $O(t)$ is defined by the time-ordered integral
\be
O(t) \equiv [e^{-i\int_{0}^{t}H_{t'} dt'}]_\mathcal{T}^{\dagger} O(0)[e^{-i\int_{0}^{t}H_{t'} dt'}]_\mathcal{T}
\ee
Note that in Eq.~\ref{Eq:Heisenberggeneral}, the Hamiltonian at time $t$ is also subject to the similarity transformation since it does not commute with the Hamiltonian at a different time.

Applying it to the boson annihilation operators, we get

\begin{align}
da_k(t)/dt &= -iE_k a_k(t) - i\lambda/E_k\cos(\Omega t)(a_k(t) + b_{-k}^{\dagger}(t))\nonumber\\
db_{-k}^{\dagger}(t)/dt &= iE_k b_{-k}^{\dagger}(t) + i\lambda/E_k\cos(\Omega t)(a_k(t) + b_{-k}^{\dagger}(t))
\label{Eq:HeisenbergFullRelativistic}
\end{align}

We focus on momenta near resonance, $\Omega \simeq 2E_k$. Without the extermal drive $a_k(t) = a_k(0)e^{-iE_k t},\  b_{-k}^\dagger(t) = b^\dagger_{-k}(0)e^{iE_k t}$. For $\lambda\ll\Omega^2$, we take the rotating wave approximation, keeping only those time-dependent terms that are in resonance,

\begin{align}
da_k(t)/dt &\simeq -iE_k a_k(t) - i\lambda/2E_k e^{-i\Omega t}b_{-k}^{\dagger}(t)\\
db_{-k}^{\dagger}(t)/dt &\simeq iE_k b_{-k}^{\dagger}(t) + i\lambda/2E_k e^{i\Omega t}a_k(t)
\end{align}
Define $\tilde{b}_k = e^{i\Omega t}b_k$, we have
\be
d\left (\begin{array}{c}
a_k(t)\\
\tilde{b}_{-k}^\dagger(t)
\end{array}\right)/dt
\simeq -i\left (\begin{array}{cc}
E_k & \frac{\lambda}{2E_k}\\
-\frac{\lambda}{2E_k} & \Omega-E_k\\
\end{array}\right)
\left (\begin{array}{c}
a_k(t)\\
\tilde{b}_{-k}^\dagger(t)
\end{array}\right)
\label{Eq:nonHermitianMatrix}
\ee

The eigenvalues of the matrix is 
\begin{align}
    \omega = \frac{\Omega}{2}\pm i\theta_k&, &\theta_k=\sqrt{(\frac{\lambda}{2E_k})^2-(\frac{\Omega}{2}-E_k)^2}
\end{align}
In the resonant region, $|\frac{\Omega}{2}-E_k|<\frac{\lambda}{2E_k}$, the eigenvalues are complex. The complex eigenvalues comes from the relative minus sign between the two off-diagonal elements in Eq.~\ref{Eq:nonHermitianMatrix}, which is a consequence of Bose statistics. If we mix two fermion bands crossing each other, we would open a gap in the energy spectrum (Fig.~\ref{Fig:drivenDispersion}(a)). These are often referred to as Flouquet bands.  However, for the boson model, the `gap opening' is in `the opposite direction'; there is a momentum range that does not have real energy eigenvalues (Fig.~\ref{Fig:drivenDispersion} (b)).

Consider turning on the drive at time $t=0$, we have

\be
\left (\begin{array}{c}
a_k(t)\\
\tilde{b}_{-k}^\dagger(t)
\end{array}\right)
\simeq e^{-i\frac{\Omega}{2}t}\left (\begin{array}{cc}
u_k(t) & v_k^*(t)\\
v_k(t) & u_k^*(t)\\
\end{array}\right)
\left (\begin{array}{c}
a_k(0)\\
b_{-k}^\dagger(0)
\end{array}\right),
\label{Eq:atbdaggert}
\ee
where $u_k(t)$ and $v_k(t)$ grows exponentially in the resonant region,
\begin{align}
u_k(t) &= \cosh(\theta_k t) + i\frac{\frac{\Omega}{2}-E_k}{\theta_k}\sinh(\theta_k t),\nonumber\\
v_k(t) &= i\frac{\lambda}{2E_k\theta_k}\sinh(\theta_k t)
\label{Eq:ukvk}
\end{align}

Eq.~\ref{Eq:atbdaggert} represents a time-dependent Bogoliubov transformation of the boson operators; $u_k(t)$ and $v_k(t)$ satisfies $|u_k(t)|^2-|v_k(t)|^2=1$. From Eq.~\ref{Eq:atbdaggert}, the number of excited bosons at time t is

\be
\<a_k^\dagger(t) a_k(t)\> =\<b_k^\dagger(t) b_k(t)\> = |v_k(t)|^2
\ee

These bosonic particles and holes exist in a coherent state. The exponentially growing mixing between $a_k$ and $b_{-k}^{\dagger}$ corresponds to an exponentially growing `particle-hole condensate' of electron pairs in the Schrodinger picture. To figure out the quantum state at time t in the Schrodinger picture, note that
\begin{align}
&(a_k(t) - \frac{v_k^*(t)}{u_k^*(t)}e^{-i\Omega t}b_{-k}^{\dagger}(t))|0\> \propto
a_k(0)|0\> = 0,\\
\Rightarrow\ 
& (a_k(0) - \frac{v_k^*(t)}{u_k^*(t)}e^{-i\Omega t}b_{-k}^{\dagger}(0))|t\>=0,
\end{align}
Similarly,
\be
(b_k(0) - \frac{v_k^*(t)}{u_k^*(t)}e^{-i\Omega t}a_{-k}^{\dagger}(0))|t\> = 0
\ee
where $|0\>$ and $|t\>$ denote the ground state and the state at time t. The two equations above completely fixes $|t\>$ up to an arbitrary overall phase $\phi_0(t)$,
\begin{align}
|t\> &= e^{i\phi_0(t)}\prod_k\sqrt{1-|v_k(t)/u_k(t)|^2}e^{\frac{v_k^*(t)}{u_k^*(t)}e^{-i\Omega t}a_k^\dagger b_{-k}^{\dagger}}|0\>\nonumber\\
&\propto e^{\sum_k\frac{v_k^*(t)}{u_k^*(t)}e^{-i\Omega t}a_k^\dagger b_{-k}^{\dagger}}|0\>.
\end{align}

For each momentum, the electron pairs and vacancies of electron pairs form a two-mode squeezed state, which is well studied in quantum optics literature. For us, it is important that excitations at different momenta are also coherent. This state represents a condensate of charge-neutral pairs of the bosons, with `wavefunction' $v^*_k(t)/u^*_k(t)e^{-i\Omega t}$ in the momentum space. 

This result is insensitive to the choice of the periodic drive. For example, if we change the periodic drive to be proportional to the current density in the x direction (for pump field in the x direction), with a coefficient $\lambda'$, we only need to change $\lambda$ into $\lambda'v^2k_x$.

At first glance, the quantum state at time $t$ represents a bosonic version of exciton condensates. However, the wavefunction indicates that the charge 2e pairs and charge -2e vacancies are not bound together; the wavefunction actually decays as a power law in real space for large t, which is reasonable since the energy of the particle and the hole lies in the continuum of a band instead of in the band gap. This property is crucial for the electromagnetic response we discuss in the next section. Under external electromagnetic field, the positive charge and the negative charge move in opposite directions; this relative motion, absent in usual exciton condensates, gives the superconducting-like response at early time.

The discussion above relies on the rotating wave approximation, which  greatly simplifies the calculation. However, it breaks gauge invariance, which is crucial for the linear response formalism, if we couple the model to electromagnetic field. To get the correct response functions of the relativistic model, we have to take into account corrections to the rotating wave approximation (see appendix~\ref{appendix:relativisticresponse}). In order to avoid this cumbersome practice whenever possible, we illustrate the physics by studying an alternative non-relativistic model, which captures the essence of the non-equilibrium physics while making the rotating-wave approximation exact.

\begin{align}
\mathcal{L}_\text{non-rel} = &\ \ ia^*\partial_t a - \frac{1}{2m}|\nabla a|^2 - \Delta |a|^2\nonumber - \frac{\tilde{U}}{4}|a|^4\\
&+ib^*\partial_t b - \frac{1}{2m}|\nabla b|^2 - \Delta |b|^2\nonumber\nonumber- \frac{\tilde{U}}{4}|b|^4\\
&-(\tilde{\lambda} e^{i\Omega t}ab + \tilde{\lambda} e^{-i\Omega t}a^*b^*),
\label{Eq:nonrelativisticLagrangian}
\end{align}
where $a$ and $b$ represents charge 2e pairs and charge \mbox{-2e} vacancies. Ignoring the interaction, the corresponding free Hamiltonian is 
\be
H_t = \sum_k\frac{k^2}{2m}(a_k^\dagger a_k+b_k^\dagger b_k) + (\tilde{\lambda} a_kb_{-k}e^{i\Omega t} + h.c.)
\label{Eq:nonrelativisticHamiltonian}
\ee 
We follow the same procedure to analyze this model. The solution to the Heisenberg equation is still in the form of Eq~\ref{Eq:atbdaggert}, except that
\begin{align}
E_k = \frac{k^2}{2m}&, &\theta_k = \sqrt{\tilde{\lambda}^2-(\frac{\Omega}{2}-E_k)^2}
\label{Eq:nonrelativisticthetak}
\end{align}
and
\begin{align}
u_k(t) &= \cosh(\theta_k t) + i\frac{\frac{\Omega}{2}-E_k}{\theta_k}\sinh(\theta_k t),\nonumber\\
v_k(t) &= i\frac{\tilde{\lambda}}{\theta_k}\sinh(\theta_k t),
\label{Eq:ukvknonrelativistic}
\end{align}
comparing with the relativistic model, we identify $\tilde{\lambda}$ approximately as $\lambda/\Omega$.

The photo-induced superconducting-like response we discuss relies on the existence of both particle and hole excitations and that they are both bosonic, but the particle-hole symmetry of the model is not important. We choose it for simplicity.

\section{dissipation}
\label{Sec: dissipation}

Before discussing the electromagnetic response of the non-equilibrium condensate , we study whether it is robust under dissipation. We distinguish two kinds of dissipation. The first kind is the scattering of a single boson, which includes scatterings by phonons and disorders to a different momentum, scatterings into two electrons, etc. The scattering rates of these processes are linearly proportional to the boson density. The second kind of dissipation is particle-hole recombination; its rate is quadratic in the boson density. We discuss it in Sec.~\ref{particleHoleRecombination}. The first kind of dissipation can be treated analytically. We use it for the calculation of conductivity in the next section.

We introduce a phenomenological decay rate $\Gamma$ per boson. We use the Lindblad equation for the evolution of density matrix $\rho$
\begin{align}
    d\rho/dt = \mathfrak{L}_t[\rho],
    \label{Eq:lidbladEq}
\end{align}
where $\mathfrak{L}$ is a linear functional acting on the space of operators,

\be
\mathfrak{L}_t[\rho] \equiv -i[H_t,\rho] + \sum_{k}\sum_{\sigma=a,b}L_{\sigma,k}\rho L^\dagger_{\sigma,k} -\frac{1}{2} \{L^\dagger_{\sigma,k} L_{\sigma,k},\rho\}.
\label{Eq:Lindbladoperator}
\ee

We choose the simplest jump operators, $L_{a,k}= \sqrt{\Gamma}a_k$, $L_{b,k} = \sqrt{\Gamma} b_k$, representing the incoherent decay of bosons at rate $\Gamma$. The formal solution of Eq.~\ref{Eq:lidbladEq} is given by the time-ordered integral
\be
\rho(t) = [ e^{\int_{0}^t \mathfrak{L}_{t'} dt' }]_\mathcal{T}\rho(0),
\ee
In the Heisenberg picture
\be
O(t) = [ e^{\int_{0}^t \mathfrak{L}_{t'} dt' }]^\dagger_\mathcal{T}O(0),
\label{Eq:LindbladHeisenberg}
\ee

The evolution of the boson creation and annihilation operators is still simple (see Appendix~\ref{appendix:dissipation} for details),
\be
\left (\begin{array}{c}
a_k(t)\\
\tilde{b}_{-k}^\dagger(t)
\end{array}\right)
\simeq e^{-i\frac{\Omega}{2}t-\Gamma t/2}\left (\begin{array}{cc}
u_k(t) & v_k^*(t)\\
v_k(t) & u_k^*(t)\\
\end{array}\right)
\left (\begin{array}{c}
a_k(0)\\
b_{-k}^\dagger(0)
\end{array}\right).
\label{Eq:atbdaggertDissipation}
\ee

Since $u_k(t)$ and $v_k(t)$ grows as $e^{\tilde{\lambda}t}$, the exponential growing condensate survives (at early time) as long as $\tilde{\lambda} > \Gamma/2$. Otherwise, the effect of the pump would be small at any time and well described by perturbative approaches.

Lastly, we point out that the dissipative evolution of operators defined by Eq.~\ref{Eq:LindbladHeisenberg} does not preserve operator multiplications (see Appendix~\ref{appendix:dissipation} for details). Specifically, we find
\be
a_k^\dagger a_k(t)\equiv [ e^{\int_{0}^t \mathfrak{L}_{t'} dt' }]^\dagger_\mathcal{T}[a_k^\dagger a_k(0)] = a_k^{\dagger}(t) a_k(t) + n^\Gamma_k(t),
\label{Eq:bosonbilinearLindblad}
\ee 
where 
\be 
n^\Gamma_k(t) \equiv \Gamma\int_0^t |v_k(t')|^2e^{-\Gamma t'} dt'.
\label{Eq:nGammak}
\ee 

\section{Conductivity at early time}
\label{Sec: linear response}

Now we study the response of the non-equilibrium condensate to a probe field at early time. We define the early time as the time that the optical phonon which drives the electron pairs has not decayed, and that the boson density is low such that the interaction of bosons are negligible. In this limit, we calculate the electromagnetic response analytically. More broadly, we shall argue that results in this section applies qualitatively as long as the boson density is growing.

We minimally couple the bosons to the electromagnetic field, $\partial_t \rightarrow \partial_t + ie^*A_0$, $\nabla \rightarrow \nabla + ie^*\vec{A}$, where $e^*=-2e$ and $(A_0,\vec{A})$ is the electromagnetic four potential. For simplicity, we first study the non-relativistic model. The electric current at momentum $\vec{q}$ is 

\begin{align}
    \vec{j}(\vec{q})_\text{non-rel} =\frac{\delta\mathcal{L}_\text{non-rel}}{\delta \vec{A}(-\vec{q})} \equiv \vec{j}^\text{P}_\text{non-rel}(\vec{q}) + \vec{j}^\text{D}_\text{non-rel}(\vec{q}),
\end{align}
where $\vec{j}^\text{P}_\text{non-rel}$ and $\vec{j}^\text{D}_\text{non-rel}$ are the paramagnetic current and the diamagnetic current,

\begin{align}
%\vec{j}^\text{P}_\text{non-rel}(\vec{q}) &=  \frac{e^*}{m}\sum_{\vec{k}}(\vec{k} + \vec{q})(-a_{\vec{k}}^\dagger a_{\vec{k}+ \vec{q}}+ b_{\vec{k}}^\dagger b_{\vec{k}+ \vec{q}}) \nonumber\\ \vec{j}^\text{D}_\text{non-rel}(\vec{q}) &= -\frac{{e^*}^2}{m}\sum_{\vec{k}} (a_{\vec{k}}^\dagger a_{\vec{k}} + b_{\vec{k}}^\dagger b_{\vec{k}})\vec{A}(\vec{q}),
\vec{j}^\text{P}_\text{non-rel}(q) &=  \frac{e^*}{m}\sum_{k}(\vec{k} + \vec{q})(-a_{k}^\dagger a_{k+ q}+ b_{k}^\dagger b_{k+ q}) \nonumber\\ \vec{j}^\text{D}_\text{non-rel}(q) &= -\frac{{e^*}^2}{m}\sum_{k} (a_{k}^\dagger a_{k} + b_{k}^\dagger b_{k})\vec{A}(q),
\label{Eq:nonrelativisticCurrent}
\end{align}
where $\vec{j}^\text{D}_\text{non-rel}$ comes from the quadratic terms of $\vec{A}$ after the substitution, $\nabla \rightarrow \nabla + ie^*\vec{A}$ into Eq.~\ref{Eq:nonrelativisticLagrangian}. To the linear order in $\vec{A}$, 

\begin{align}
   \vec{j}^\text{D}_\text{non-rel}(q) &= -\frac{{e^*}^2}{m}\sum_{k} \<a_{k}^\dagger a_{k}(t) + b_{k}^\dagger b_{k}(t)\>\vec{A}(q),\nonumber\\ 
   &= -2\frac{{e^*}^2}{m}\sum_{k}(|v_{k}(t)|^2 + n^\Gamma_{k}(t)) \vec{A}(q)\nonumber\\
   &\simeq -\frac{{e^*}^2}{8}\frac{\tilde{\lambda}}{\tilde{\lambda}-\Gamma/2}\sqrt{\frac{\tilde{\lambda}}{\pi t}}e^{2\tilde{\lambda} t - \Gamma t} \vec{A}(q),
   \label{Eq:diamagneticcurrentnonrelativistic}
\end{align}
where we have used Eq.~\ref{Eq:ukvknonrelativistic},~\ref{Eq:atbdaggertDissipation}, and Eq.~\ref{Eq:bosonbilinearLindblad}-\ref{Eq:nGammak}, and we have restricted the model to two-dimensional space. In the last line, we make the approximation $t\gg 1/\tilde{\lambda}$ and expand the growth exponent $\theta_k$ to the second order in $(\Omega/2 - E_k)$, so that we can approximate the momentum summation by a Gaussian integral. Apart from this approximation, Eq.~\ref{Eq:diamagneticcurrentnonrelativistic} holds only at early time when the boson interaction is negligible. At later time, the exponential growth saturates and decays due the boson interaction and dissipation.

This diamagnetic current resembles the London equation in a superconductor, with a superfluid density
$\rho_s = \frac{1}{8}\frac{\tilde{\lambda}}{\tilde{\lambda}-\Gamma/2}\sqrt{\frac{\tilde{\lambda}}{\pi t}}e^{2\tilde{\lambda} t - \Gamma t}$ at early time,
giving a conductivity $\sigma(\omega) = {e^*}^2\rho_s(i/\omega + \delta(\omega))$, but we have to take into account the paramagnetic current. In fact, in a free fermion system, the paramagnetic current cancels the singular superconducting response, and changes the ac conductivity into a Drude peak of width $\Gamma$.

Surprisingly, we find that, the paramagnetic current is much smaller than the diamagnetic current, even with dissipation, as long as $1/t\ll \omega,\tilde{\lambda}$. In other words, as long as the time is long enough to resolve the frequency and the growth exponent, the ac conductivity resembles that of a superconductor.

Now we sketch the calculation of the paramagnetic current without dissipation. The calculation with dissipation is very similar except that one has to deal with operator multiplications more carefully, which we discuss in Appendix~\ref{appendix:dissipation}.

Without dissipation, the conventional linear response formalism holds. But in our case the response depends not only on the frequency of the probe field but also on the time relative to the pump. We consider turning on the probe field of frequency $\omega$ slowly, \textit{before} the pump, and calculate the current at time $t$ after the pump. For large frequencies, $\omega\gg \tilde{\lambda},1/t$, the response at time $t$ mainly comes from a time window of width $\sim 1/\omega$ before time $t$; therefore it does not matter whether the probe field is turned on before the pump or after the pump. Nonetheless, this setup has the advantage of studying the crossover from large frequencies to small frequencies. 

For $\vec{q}=0$,

\begin{align}
\<j_l^P(t)\> &= \int_{-\infty}^{t}
i\<[j_l^P(t),j_m^P(t')]\>A_m(\omega)e^{-i\omega t}\nonumber\\
&=-2 \int_{-\infty}^{t}\text{Im}[\<j_l^P(t) j_m^P(t')\>]A_m(\omega)e^{-i\omega t}.
\label{Eq:linearresponseformalism}
\end{align}

Using the definition of the current in Eq.~\ref{Eq:nonrelativisticCurrent} and the evolution of the boson annihilation operators given by Eq.~\ref{Eq:atbdaggert} and Eq.~\ref{Eq:ukvknonrelativistic}, we find

\be 
\<j_x^P(t) j_x^P(t')\> = \sum_k \frac{4{e^*}^2}{m^2}k_x^2u_k(t)v_k(t)u^*_k(t')v^*_k(t')
\ee
In the limit $\omega t\gg 1, \tilde{\lambda}t\gg 1$, the summation is dominated by those momenta satisfying $\Omega/2 =E_k$, which have the largest growth exponent. However, $\<j_x^P(t) j_x^P(t')\>$ is purely real for those momenta; therefore the leading-order contribution to the response kernel is zero according to Eq.~\ref{Eq:linearresponseformalism}. The subleading contribution is smaller by at least a factor of $1/(\tilde{\lambda}t)$ or $1/(\omega t)$ compared to the diamagnetic current. In Appendix~\ref{appendix:dissipation}, we show that the same argument holds even with dissipation. Thus we confirm that in the limit $1/t\ll \omega,\tilde{\lambda}$, subjecting to the condition that the boson density is small enough such that the boson interaction is negligible, there is a superconducting-like response:

\begin{align}
    &\sigma(\omega) = i{e^*}^2\rho_s/\omega, &\rho_s = \frac{1}{8}\frac{\tilde{\lambda}}{\tilde{\lambda}-\Gamma/2}\sqrt{\frac{\tilde{\lambda}}{\pi t}}e^{2\tilde{\lambda} t - \Gamma t}
\end{align}
The exponential growth of the `superfluid density' eventually saturates due to interaction.

Note that for $\omega= 0$, the probe field is pure gauge, and gauge invariance requires that $\vec{j}\rightarrow 0$. Assuming the response is a smooth function of $t$, we should have $\vec{j}\rightarrow 0$ in the limit $\omega t\ll 1$.

The response of the relativistic model is more complicated. The paramagnetic current is no longer negligible even in the limit $\omega t \gg 1$. We show in Appendix~\ref{appendix:relativisticresponse} that in the absence of dissipation, there is a similar superconducting-like ac response,

\be 
\rho_s = \frac{1}{8}(1+4\Delta^2/\Omega^2)\sqrt{\frac{\lambda}{\pi \Omega t}}e^{2\lambda t/\Omega}
\ee 
We expect this response to be robust against dissipation as in the non-relativistic model. In addition, there is a fast oscillating current at frequency $\omega\pm \Omega$ in the relativistic model, which we ignore for now.

Physically, we have a charge-neutral, particle-hole condensate, located on a ring satisfying $E_k \simeq \Omega/2$ in the B.Z., different from a conventional superconductor, which is a condensate of charged particles at a single momentum. The reason that the sample conducts perfectly despite static scattering and dissipation in the early time is also different from that of a superconductor. 

We give a physical argument for the perfect-conductivity of the driven boson system through the comparison to a Fermi liquid. In a Fermi liquid, the acceleration of the electrons by the electric field is constantly offset by elastic scatterings with disorders and inelastic scatterings with other excitations. For the driven boson system we study, such scatterings also reduce the current. However, the imbalance of the particle and hole distribution created by the electric field cannot be erased by a single scattering event. On the contrary, it is enhanced by the periodic drive due to the Bose statistics. Thus, as long as the scattering rate is smaller than the growth rate of the condensate, we have a perfect conductor.

An interesting question is whether the transient state exhibits Meissner effect. Consider turning on a static magnetic field before the pump, would it be repelled out of the sample after the pump? In the next section, we give a physical argument that there is no Meissner effect at early time. In Appendix~\ref{appendix:Meissner}, we calculate the response to the vector potential $A(\omega,\vec{q})$ in the limit $\omega t\gg 1, \sqrt{\Omega/m}qt\gg 1$. We find that for the non-relativistic model, the results of this section hold for $\omega\gg \frac{\Omega}{\lambda}\frac{q^2}{2m}$. But for $\omega\ll \frac{\Omega}{\lambda}\frac{q^2}{2m}$, the paramagnetic current exactly cancels the diamagnetic current at small frequencies and momenta. Thus, there is no Meissner effect for strictly static magnetic field, but we may still see a perfect diamagnetism if the Meissner effect is probed by an ac field.

\section{Absence of Meissner effect at early time}

\begin{figure}[htb]
\begin{center}
\includegraphics[width=0.4\textwidth]{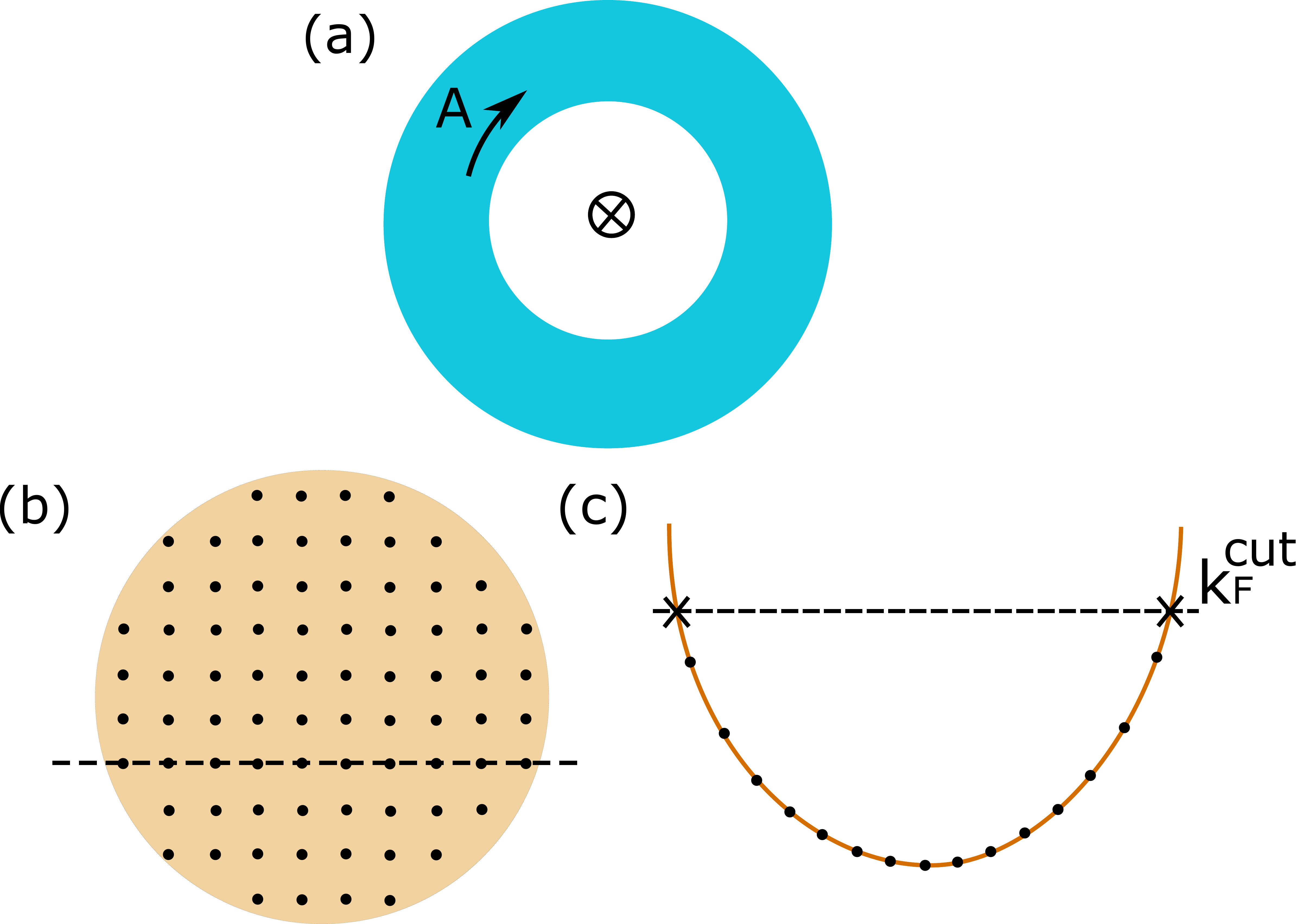}
\caption{(a) Corbino geometry. Threading a flux (the cross in the middle) through the sample (the blue ring) (b) Illustration of the quantized momenta, $\frac{2\pi}{L}(n_x,n_y)$, in a Fermi surface. The dashed line represents a cut in the B.Z. passing through the Fermi surface (c) Dispersion along the cut passing through the Fermi surface}
\label{Fig:Meissner}
\end{center}
\end{figure}

The absence of Meissner effect is shown by explicitly calculating the response function to a transverse electromagnetic potential in the static limit for the non-relativsitic model, which we do in Appendix~\ref{appendix:Meissner}. Here we present a physical argument for the absence of Meissner effect at early time. 
To see this we first review how a free boson condensate and a fermi sea respond differently  to a static transverse vector potential. Consider the Corbino geometry shown in Fig.~\ref{Fig:Meissner}. Thread a flux in the middle, which provides a vector potential $A<< 2\pi/L_x$, where $L_x$ is the circumference of the sample and we have set the charge to 1. This flux is equivalent to a twist of the boundary condition, hence the change of momentum quantization from $(2\pi n_x/L_x, 2\pi n_y/L_y)$, where $n_x$ and $n_y$ are integers, to $(2\pi n_x/L_x + A, 2\pi n_y/L_y)$. For a Bose-Einstein condensate (BEC), the single-particle momentum at which the bosons condense is forced to move from zero to $A$, which gives a current $j = NA/m$, where $N$ is the number of bosons in the condensate. This macroscopic coefficient indicates a large energy penalty for a flux, which leads to flux quantization in the Corbino geometry, a manifestation of the Meissner effect.

For a free fermion system (Fig.~\ref{Fig:Meissner}(b)), the change of the quantization condition would also give a current if the occupation number at each momentum were unchanged. This current corresponds to the diamagnetic current in the linear response formalism. However, in the static limit, the system will lower its energy by redistributing the fermion occupation, so that the fermions at some momenta are moved outside the Fermi surface and the opposite momenta are moved inside. In the new ground state, these fermion occupation relaxes to the quantized momenta which have moved into the Fermi surface under the twist of the boundary condition, reducing the total energy cost and the current. 

To see this in more detail, we consider each one-dimensional cut passing across the Fermi sea (Fig.~\ref{Fig:Meissner}(c)), the change of the momentum quantization directly adds $N^\text{cut} A$ to the total momentum where $N^\text{cut}$ denotes the total number of occupied states along this cut. In addition, if $k^\text{cut}_\text{F}$ lies in a window of width $A$ in the $2\pi/L_x$ interval between neighboring momenta, the rightmost state moves out of the Fermi surface. Removing it reduces the total momentum by $k^\text{cut}_\text{F}$. Similarly, the leftmost momentum moves inside the Fermi surface. Adding it reduces the total momentum by $k_\text{F}^\text{cut}$. On average, the change of total momentum along the cut is
\be 
\Delta k = N^\text{cut}A - k^\text{cut}_\text{F}\frac{A}{2\pi/L_x} - k^\text{cut}_\text{F}\frac{A}{2\pi/L_x} = 0.
\ee 
For spatial dimensions higher than one, averaging over all cuts, the overall change of the total momentum is zero. Thus, on average, the Fermi surface is not shifted by the vector potential. The energy gain is zero up to and including order $A^2$ and there is no current to linear order in $A$. In the case of a superconductor, a gap appears at the Fermi level, and this kind of re-distribution of occupation cannot occur.  The ground state energy increases as $A^2$, just like the the case of the boson condensate.

For the driven non-relativistic boson model we consider, the situation is closer to the fermi sea than the boson condensate. This is because the particles in the charge-neutral condensate are concentrated on a ring in momentum space satisfying

\be 
E^{p}_k + E^{h}_{-k} \simeq \Omega/2, 
\ee
where $E^{p}_k = k^2/2m^p$ and $E^{h}_k = k^2/2m^h$ are the particle and hole dispersions, which in general have different masses. As pointed out in Fig. \ref{Fig:drivenDispersion}, there is no energy gap, but a resonant ring in k space  whose occupation grows with time. At  time $t$ the effective width of the resonant ring is $\delta E\sim\sqrt{\tilde{\lambda}/t}$. Applying a small vector potential, the momentum of the particle/hole is shifted by $\pm A$. The resonant condition changes to 
\begin{align}
&\frac{(k+A)^2}{2m^p} + \frac{(-k-A)^2}{2m^h} \simeq \Omega /2\nonumber\\
\Leftrightarrow &\frac{(k+A)^2}{2m^\text{eff}} \simeq\Omega/2,
\end{align}
where $ \frac{1}{m^\text{eff}} = \frac{1}{m^p} +\frac{1}{m^h}$.
This is the same condition as for the Fermi surface of the free Fermi sea. Allowing for redistribution of the states that satisfy the resonance condition, we conclude that on average, the resonant ring is not shifted by the vector potential up to and including order $A^2$, for the same reason that the Fermi surface is not shifted by the vector potential. 

Thus, the driven boson system does not generate a large diamagnetic current that can repel a magnetic flux that is turned on before the pump. This is very different from the response to an ac field, where the electric field moves the particles; the imbalance the field creates, the current, is magnified by the parametric amplification despite the dissipation. We have an example of a perfect conductor with no Meissner effect!

\section{Late-time behavior: boson interaction, dissipation, and the decay of the phonon}
\label{Sec: late time behavior}

\begin{figure*}[htb]
\begin{center}
\includegraphics[width=0.8\textwidth]{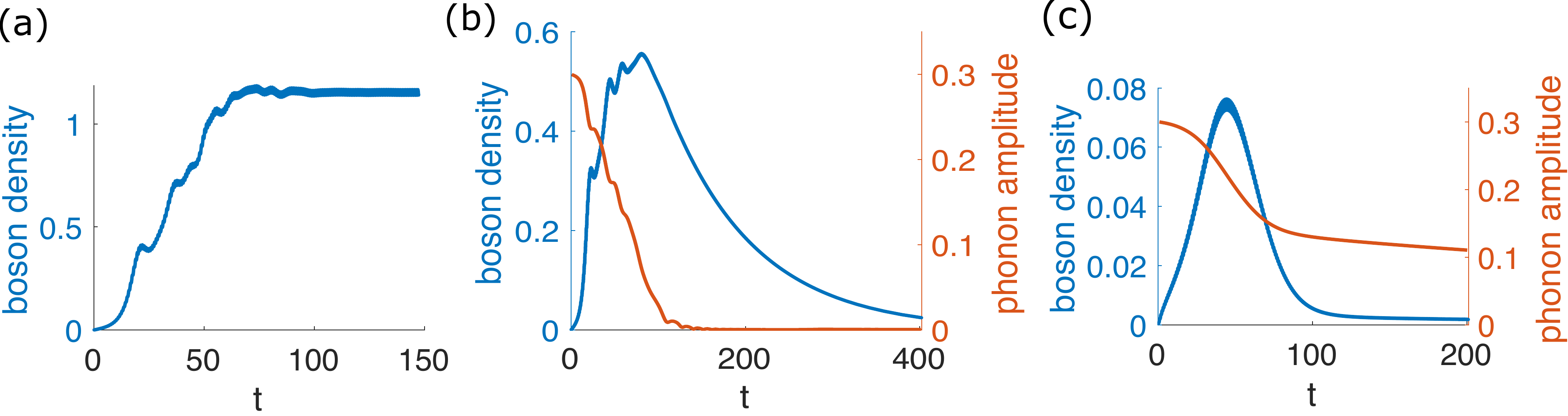}
\caption{Self-consistent solution of the non-relativistic boson and the phonon in three different scenarios. $\Omega=4,\Delta=1,m=0.2, \tilde{\lambda}_0=0.15$ for all figures (a) $ g\rightarrow 0, \Gamma\rightarrow 0$. The boson density saturates due to the interaction. The phonon amplitude stays constant. (b) $g=0.1, \Gamma=0.01$. (c) $g=0.1, \Gamma=0.2$.}
\label{Fig:latetimeNonrelativistic}
\end{center}
\end{figure*}

In the previous few sections, We discuss the early-time behavior of the driven boson system, ignoring the boson interaction and the decay of the periodic drive. In this section, we incorporate these two factors in a mean field treatment, and discuss the late-time behavior. For the decay of the periodic drive, we focus on the case of YBCO, where an optical phonon, initially excited by the short THz pump, drives the electron pairs and gradually decays as its energy transfers to the electron pairs. The electromagnetic pulse itself also drives the Cooper pairs directly, but it is usually less important in experiments conducted on YBCO since the duration of the pulse is much shorter than the lifetime of the phonon. For simplicity, we shall mainly focus on the non-relativistic boson model. We consider the following Lagrangian

\begin{align}
\mathcal{L}_\text{non-rel} = &\ \ ia^*\partial_t a - \frac{1}{2m}|\nabla a|^2 - \Delta |a|^2\nonumber - \frac{\tilde{U}}{4}|a|^4\\
&+ib^*\partial_t b - \frac{1}{2m}|\nabla b|^2 - \Delta |b|^2\nonumber\nonumber- \frac{\tilde{U}}{4}|b|^4\\
&-2\tilde{\lambda} (ab + a^*b^*) +\frac{1}{g}(\partial_t \tilde{\lambda})^2 - \frac{\Omega^2}{g}\tilde{\lambda}^2,
\label{Eq:nonrelativisticphononLagrangian}
\end{align}
where $\tilde{\lambda}$ is the amplitude of the optical phonon (the c-aixs displacement of the apical oxygen shown in Fig.~\ref{Fig:YBCO}, scaled to remove the coupling constant of the phonon and the boson), oscillating at frequency $\Omega$. Its equation of motion is

\be 
\partial_t^2\tilde{\lambda}(x) = -\Omega^2 \tilde{\lambda}(x) - g [a(x) b(x) + a^\dagger(x) b^\dagger(x)]
\ee 
Since the phonon amplitude, initially created by the pump, is spatially uniform and much larger than the quantum zero-point motion and it couples to all boson modes, we expect the following classical approximation to be appropriate.

\be 
\partial_t^2\tilde{\lambda} = -\Omega^2 \tilde{\lambda} - \frac{2g}{N}\sum_k \text{Re}[\<a_k b_{-k}\>].
\label{Eq:phonondecay}
\ee
$N$ is the total number of discrete momenta. 

We also approximate the interaction $\tilde{U}|a|^4$ ($\tilde{U}|b|^4$) by $4\tilde{U}\<|a|^2\>|a|^2$ ($4\tilde{U}\<|b|^2\>|b|^2$), where the factor of 4 comes from different ways of contracting the fields. This gives the mean field dispersion

\be 
E_k^\text{MF} = E_k + \frac{\tilde{U}}{N}\sum_k\<a^\dagger_k a_k\>
\label{Eq:meanfielddispersion}
\ee 

Now we discuss the scenario that the decay of the phonon and the dissipation of the boson are negligible, $g\rightarrow 0, \Gamma\rightarrow 0$. In this scenario, the late-time behavior is dictated by the interaction, which we incorporate into the Heisenberg equation by changing the bare dispersion to the mean field dispersion. 

Immediately after the pump, The boson number grows exponentially at resonant momenta, as discussed before. As it grows, the mean field energy also goes up, and the resonant condition changes. The resonant region becomes considerably shifted when $\tilde{U}\<|a|^2\>\sim \tilde{\lambda}$. Eventually, when $\tilde{U}\<|a|^2\>+\Delta =\Omega/2 +\tilde{\lambda}$, no boson modes are resonantly excited and the growth saturates (Fig.~\ref{Fig:latetimeNonrelativistic}(a)). Since $\tilde{\lambda}<< \Omega$, the boson density at saturation, $\<|a|^2\>= (\Omega/2 +\tilde{\lambda} - \Delta)/U$, is almost independent of the amplitude of the pump. 

If the periodic drive persists long enough, the simplest possibility is that the excited bosons relax to the zero momentum and condense at the frequency $\Omega/2$. According to Eq.~\ref{Eq:nonrelativisticthetak} and \ref{Eq:meanfielddispersion}, when the boson density saturates, the effective dispersion of the driven system is (see Fig.~\ref{Fig:drivenDispersion}(c))

\begin{align}
   \omega &= \Omega/2 \pm \sqrt{\tilde{\lambda}^2-(\frac{\Omega}{2}-E^\text{MF}_k)^2}\nonumber\\
   &= \Omega/2 \pm |k|\sqrt{\tilde{\lambda}/m + \frac{k^2}{4m^2}} 
   \label{Eqsteadystatedisersion}
\end{align}
The `soft mode' near $\Omega/2$ has linear dispersion, just like the Goldstone mode in an equilibrium superconductor! In fact, this non-equilibrium system holds rich phenomena of novel superconducting steady states, some of which the mean field theory is inadequate to describe. We analyze the phase diagram of the steady states in a separate work~\cite{steadyState}.

\begin{figure}[htb]
\begin{center}
\includegraphics[width=0.5\textwidth]{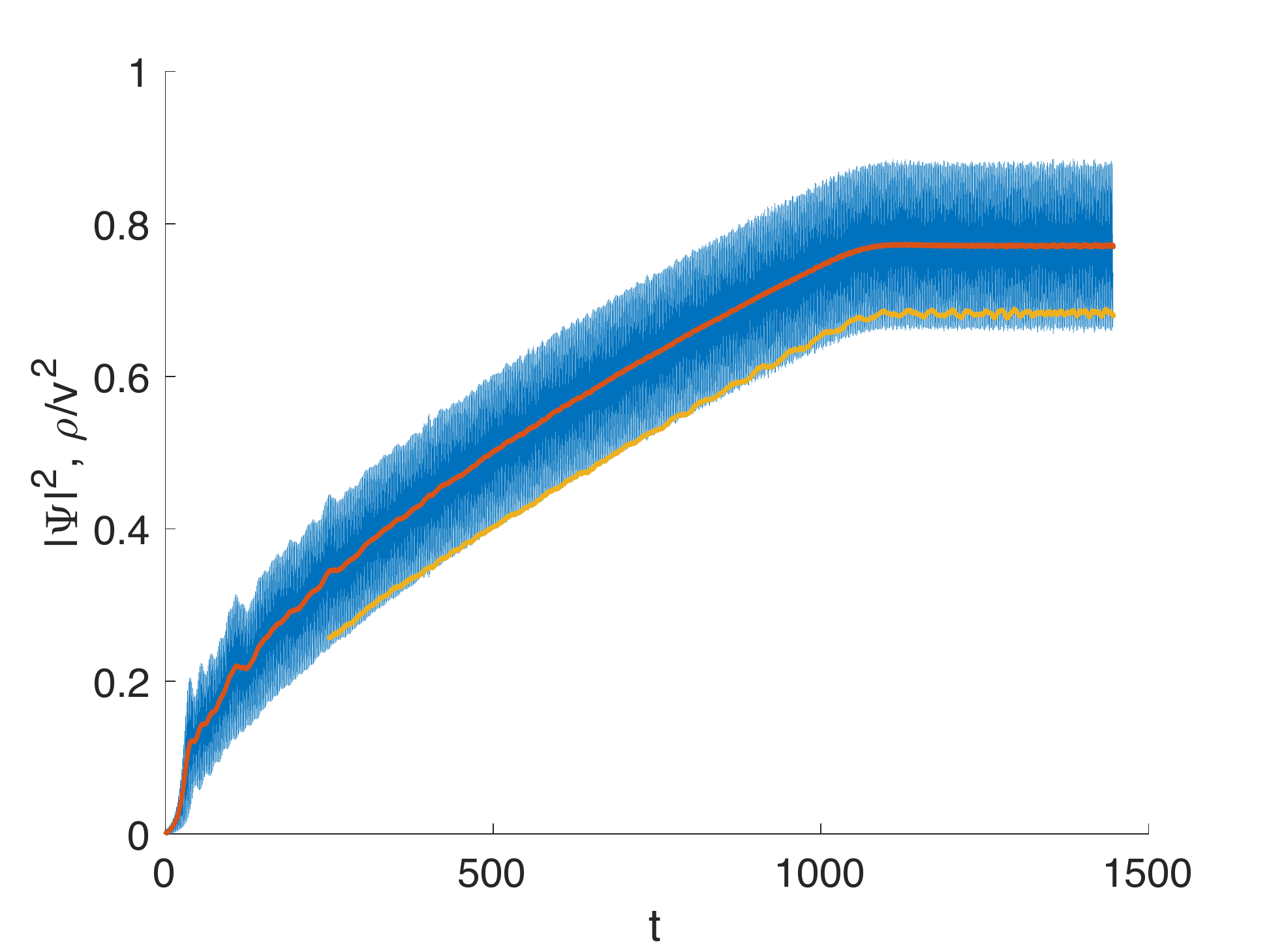}
\caption{Boson amplitude and `superfluid density' of the relativistic model as a function of time. The blue curve represents $|\Psi^2(t)|$, which has a slowly-varying component and an oscillating component near the frequency $\Omega=4$. The red curve represents the slowly-varying component of $|\Psi^2(t)|$. The yellow curve represents the response coefficient to an ac probe field with frequency $\omega =0.5$, which is slowly turned on after the pump in a time window of $\Delta t\sim 150$.}
\label{Fig:latetimeRelativistic}
\end{center}
\end{figure}

We also simulate the relativistic boson model. In addition to the smooth growth and saturation seen in the non-relativistic model, $\<|\Psi|^2\>$ also has a frequency-$\Omega$ oscillating component. Furthermore, we compute the `superfluid density' numerically at a low probe frequency (Fig.~\ref{Fig:latetimeRelativistic})\footnote{In the computation of the response function, we ignore the vertex correction, which is at the same order of magnitude as the mean field energy modification, but hard to deal with in this time-dependent model.}. We use a lattice regularization for the relativistic dispersion, which is discussed in Appendix.~\ref{appendix:relativisticresponse}.

For the majority of the experiments, the energy absorbed from the pump is not enough for the bosons to reach the steady state. In the latest result on YBCO, it is shown that the extrapolated superfluid density is proportional to the electric field of the pump (Fig.S7 in Ref.~\cite{PhysRevX.10.011053}). Within our model, this experimental result implies the growth of the boson density stops before reaching the steady state due to the decay of the optical phonon and the dissipation of the boson. In the rest of this section, we first include the scattering of a single boson, where we give a self-consistent mean field solution. Then we discuss the case where particle-hole recombination is the main source of dissipation.

\subsection{Scattering of a single boson}

Combining the mean-field treatment discussed above and the Lindblad equation discussed in Sec.~\ref{Sec: dissipation}, we derive the following equations for the average values of the boson bilinears (see Appendix~\ref{appendix:dissipation} for details)

\begin{align}
    \partial_t\<a_k^\dagger a_k\> &= -\Gamma \<a_k^\dagger a_k\> - 4\tilde{\lambda}\ \text{Im}[\<a_kb_{-k}\>]\nonumber\\
    \partial_t \<a_k b_{-k}\> &= (-2iE_k^\text{MF} -\Gamma)\<a_k b_{-k}\> -2i\tilde{\lambda}(2\<a_k^\dagger a_k\> + 1)
    \label{Eq:MeanfieldEqExpectation}
\end{align}
We choose the initial condition $\<a_k^\dagger a_k\>=\<a_k b_{-k}\> = 0, \tilde{\lambda}=\lambda_0, \partial_t\tilde{\lambda}=0$ and solve Eq.~[\ref{Eq:phonondecay}, \ref{Eq:meanfielddispersion}, \ref{Eq:MeanfieldEqExpectation}] numerically. In the following, we discuss two scenarios.

In the first scenario, the dissipation rate $\Gamma$ is small but nonzero, and the phonon susceptibility $g$ is chosen so that the total energy initially stored in the phonon is comparable to the total energy of the boson system if it were to reach the steady state. In this scenario, the boson density initially grows as in the first scenario, but as the phonon energy transfers to the boson system, the growth slows down and eventually reverses as the energy dissipates away (Fig.~\ref{Fig:latetimeNonrelativistic}(b)).

In the second scenario, the dissipation rate is larger than that in the second scenario, and is comparable to the initial growth rate of the boson. Once the phonon amplitude is not large enough to compensate the dissipation, the boson density starts to decay sharply. Meanwhile the phonon itself still carries considerable energy which decays much more slowly (Fig.~\ref{Fig:latetimeNonrelativistic}(c)).

However, none of these two scenarios produce the linear proportionality between the superfluid density and the electric field seen in experiments. If the dissipation rate is comparable to the initial growth rate as in the second scenario, we should observe that the superfluid density vanishes at a nonzero field. On the other hand, if the dissipation rate is much smaller than the initial growth rate as in the first scenario, we expect the boson density, which is proportional to the superfluid density, to be proportional to the total energy of the pump, which is quadratic in the electric field. In order to resolve this problem, we next discuss the second kind of dissipation mentioned earlier, namely that due to recombination of particle and hole.

\subsection{particle-hole recombination}

In fact, the linear proportionality between the superfluid density and the electric field is natural if the main source of dissipation is the recombination of the bosonic particle and hole. Phenomenologically, we can add a decay rate proportional to the square of the boson density in the initial exponential-growing regime

\be 
\frac{d\rho}{dt} = 2\tilde{\lambda}\rho - \Gamma'\rho^2,
\ee 
where $\rho = \<|a|^2\> = \<|b|^2\>$. Then the maximum boson density is $\rho_\text{max}=2\tilde{\lambda}/\Gamma'$, which is linearly proportional to the electric field. 

\label{particleHoleRecombination}

\section{Summary and Outlook}

In this work we provide an explanation of the recently discovered photo-induced superconducting-like response in YBCO %and $\K3C60$ 
using a model of driven gapped electron pairs. The mechanism does not rely on other assumptions except that there are gapped electron pairs, which we previously proposed to explain the pseudogap phenomenology in cuprates~\cite{PhysRevB.101.064502}. In the early time after the pump, when the interaction of the excited electron pairs and vacancies of pairs are negligible, we show that the non-equilibrium state is a time-dependent particle-hole condensate, which exhibits a growing perfect-conductor response even with dissipation. In the late time the condensate saturates and decays. Our observations support the existence of gapped electron pairs in the pseudogap region. It would be very interesting to further investigate the transient state by other experimental techniques, for example, time-dependent ARPES, and tunneling measurements, and make comparisons with the theory. 

In the following we discuss several open problems we would like to study in the future.

We have modeled disorders and dissipation effectively as the decay of the of the excited electron pairs and vacancies of pairs, like in the Drude model. We have shown that the superconducting-like response is robust against this decay. However, it would be good to explicitly include quenched disorders in the model. Following Anderson's argument~\cite{anderson1959theory} on disordered superconductors in equilibrium, we find that the coherent particle-hole condensate is stable against disorders: instead of using the momentum eigenstates for electron pairs, we can solve the Heisenberg equation of motion in the basis of the energy eigenstates with quenched disorders. This approach preserves the two by two block structure and the analytic solution (Eq.~\ref{Eq:atbdaggert}) holds in the new basis. Yet, the response of this condensate to electromagnetic fields requires further investigation.

The present work focus on the in-plane response of the 2D system. In a follow-up work~\cite{bilayer}, we generalize our model to coupled bi-layer systems and discuss the c-axis response.

We have shown that at least in the early time, the system shows superconducting-like ac conductivity but no Meissner effect for static magnetic field. It is related to that electron pairs condense in a ring of momenta satisfying $E_k =\Omega/2$ instead of at a single momentum. Can such a state exist in equilibrium? Would it be a superconductor? It would also be interesting to further explore the possibility of a perfect conductor without Meissner effect in non-equilibrium and equilibrium systems.

Our mean-field treatment of the late-time behavior is adequate for the majority of the experiments, which uses a short pump (around 100fs). But it is not accurate enough to resolve the fine structures of the potential steady states, which may describe the most recent experiment on $\K3C60$~\cite{budden2020evidence,steadyState}, where the pump lasts for a few picoseconds and the superconducting-like response survives much longer. This fascinating phenomena deserves further investigation.

\section{Acknowledgments}

We are grateful to Andrea Cavalleri and Eugene Demler for useful discussions and correspondence. This research is funded in part by the Gordon and Betty Moore Foundation. P.A.L. acknowledges the support by DOE office
of Basic Sciences Grant No. DE-FG02-03ER46076.

\bibliographystyle{apsrev4-1}
\bibliography{photoSC.bib}

\appendix
\section{Dissipation}
\label{appendix:dissipation}

In this appendix we discuss the mathematical treatment of dissipation. We first derive the time evolution of boson operators $a_k$ and $b_k$ and boson bilinears, and then discuss the linear response formalism and specifically, the paramagnetic current with dissipation.

\subsection{Evolution of boson operators}

According to Eq.~\ref{Eq:LindbladHeisenberg}, operators in the Heisenberg picture is defined as
\be 
O(t) = e^{\mathfrak{L}^\dagger_{0} \Delta t} e^{\mathfrak{L}^\dagger_{\Delta t} \Delta t}\cdots e^{\mathfrak{L}^\dagger_{t} \Delta t}O(0),
\ee 
where the operation $\mathfrak{L}^\dagger_t$ is defined by the equation
\be 
\text{Tr}[O\mathfrak{L}_t[\rho]] = \text{Tr}[\mathfrak{L}_t^\dagger[O]\rho]
\ee 
Thus
\be
\mathfrak{L}^\dagger_t[O] \equiv i[H_t,O] + \sum_{k}\sum_{\sigma=a,b}L^\dagger_{\sigma,k}O L_{\sigma,k} -\frac{1}{2} \{L^\dagger_{\sigma,k} L_{\sigma,k},O\}.
\label{Eq:Lindbladoperator}
\ee

For the free non-relativistic boson model,
\begin{align}
\left(\begin{array}{c}
\mathfrak{L}_t^\dagger[a_k]\\
\mathfrak{L}_t^\dagger[b_{-k}^\dagger]
\end{array}\right)
&= \left(\begin{array}{cc}
-iE_k-\Gamma/2 & -i\tilde{\lambda} e^{-i\Omega t}\\
i\tilde{\lambda} e^{i\Omega t} & iE_k -\Gamma/2
\end{array}\right)
\left(\begin{array}{c}
a_k\\ b_{-k}^\dagger
\end{array}\right)\nonumber \\
&\equiv D_t \left(\begin{array}{c}
a_k\\ b_{-k}^\dagger
\end{array}\right)
\end{align}
Thus,
\begin{align}
\left(\begin{array}{c}
a_k(t)\\ b_{-k}^\dagger(t)
\end{array}\right)
&=e^{\mathfrak{L}^\dagger_{0} \Delta t}\cdots e^{\mathfrak{L}^\dagger_{t-\Delta t} \Delta t} e^{\mathfrak{L}^\dagger_{t} \Delta t}
\left(\begin{array}{c}
a_k(0)\\ b_{-k}^\dagger(0)
\end{array}\right)\nonumber\\
&=e^{\mathfrak{L}^\dagger_{0} \Delta t}\cdots e^{\mathfrak{L}^\dagger_{t-\Delta t} \Delta t} e^{D_{t} \Delta t}
\left(\begin{array}{c}
a_k(0)\\ b_{-k}^\dagger(0)
\end{array}\right)\nonumber\\
&=e^{D_{t} \Delta t}e^{\mathfrak{L}^\dagger_{0} \Delta t}\cdots e^{\mathfrak{L}^\dagger_{t-\Delta t} \Delta t}
\left(\begin{array}{c}
a_k(0)\\ b_{-k}^\dagger(0)
\end{array}\right)\nonumber\\
&=e^{D_{t}\Delta t}\cdots e^{D_{0}\Delta t}
\left(\begin{array}{c}
a_k(0)\\ b_{-k}^\dagger(0)
\end{array}\right)
\end{align}
Note that $\mathfrak{L}_t^\dagger$ does not act on c numbers, and because of that the last line has the opposite time order as the first line. We now get a simple equation of the time evolution
\be
d\left (\begin{array}{c}
a_k(t)\\
b_{-k}^\dagger(t)
\end{array}\right)/dt
\simeq \left (\begin{array}{cc}
-iE_k-\Gamma/2 & -i\tilde{\lambda} e^{-i\Omega t}\\
i\tilde{\lambda} e^{i\Omega t} & iE_k - \Gamma/2\\
\end{array}\right)
\left (\begin{array}{c}
a_k(t)\\
b_{-k}^\dagger(t)
\end{array}\right),
\ee
which leads to Eq.~\ref{Eq:atbdaggertDissipation} in the main text.

Unlike unitary evolution, the Lindblad equation does not preserve operator multiplications. For example,
\be 
\mathfrak{L}_t^\dagger[a_k^\dagger a_k] = \mathfrak{L}_t^\dagger[a_k^\dagger] a_k + a_k^\dagger \mathfrak{L}_t^\dagger[a_k],
\ee 
but
\be 
\mathfrak{L}_t^\dagger[a_k a_k^\dagger] = a_k\mathfrak{L}_t^\dagger[a_k^\dagger]  +  \mathfrak{L}_t^\dagger[a_k]a_k^\dagger + \Gamma.
\ee

A simple way to solve the time evolution of boson bilinear operators is to first remove the time dependence of the Hamiltonian (Eq.~\ref{Eq:nonrelativisticHamiltonian}) by a unitary transformation
\be 
U(t) = e^{i\sum_k b_k^\dagger b_k \Omega t},
\ee 
such that
\be 
\frac{d}{dt} a_k^\dagger a_k(t) = \mathfrak{L}^\dagger [a_k^\dagger a_k(t)]
\label{Eq:appendixLindbladHeisenbergtimeindependent}
\ee 
Assuming $a_k^\dagger a_k(t) = a_k^\dagger(t) a_k(t) + n^\Gamma_k(t)$, using the solution that $a_k(t) = e^{-i\Omega t/2-\Gamma t/2} (u_k(t)a_k + v_k(t)b_{-k}^\dagger)$, and comparing the left hand side of Eq.~\ref{Eq:appendixLindbladHeisenbergtimeindependent} with the right hand side, we find that
\be 
\frac{d}{dt}n^\Gamma_k(t) = \Gamma|v_k(t)|^2e^{-\Gamma t},
\ee 
which gives Eq.~\ref{Eq:nGammak}. Similarly,
\begin{align}
a_k b_{-k}(t) &= a_k(t)b_{-k}(t) + m^\Gamma_k(t)\\
\frac{d}{dt} m^\Gamma_k(t) &= \Gamma u_k(t')v^*_k(t')e^{-\Gamma t'}
\label{mkGamma}
\end{align}

Lastly, for the joint boson-phonon evolution discussed in Sec.~\ref{Sec: late time behavior}, we cannot remove the time dependence of the mean-field Hamiltonian by a unitary transformation. Instead, we use the following equation of motion of the average value of an operator

\begin{align}
\frac{d}{dt} \< O\>= \text{Tr}[O\mathfrak{L}_t[\rho(t)]] = \text{Tr}[\mathfrak{L}_t^\dagger[O]\rho(t)] = \< \mathfrak{L}_t^\dagger[O]\>,
\end{align}
which gives Eq.~\ref{Eq:MeanfieldEqExpectation}.

\subsection{Linear response with dissipation}

Now we discuss modifications of the linear response formalism in the presence of dissipation. We show that for the non-relativistic boson model the paramagnetic current is much smaller than the diamagnetic current in the limit $1/t\ll \omega,\tilde{\lambda}$. 

For simplicity, we first remove the time-dependence of the periodic Hamiltonian by a unitary transformation and then formulate the linear response theory in the new basis.

Applying a probe vector potential in the x direction, the Hamiltonian is modified as
\be 
H_A = H - j_xA_x(t)
\ee 
In the Schrodinger picture, to the first order in $A_x$
\be
\frac{d}{dt}\rho(t) = \mathfrak{L}[\rho(t)] + i[j_x^P,\rho(t)]A_x(t),
\ee 
\begin{align}
\delta\rho(t) &= i\int_{-\infty}^{t}e^{\mathfrak{L}(t-t')}\{[j_x^P,\rho(t')]\}A_x(t')dt'\\
&=i\int_{-\infty}^{t}e^{\mathfrak{L}(t-t')}\{j_x^P e^{\mathfrak{L}t'}[\rho(0)] - e^{\mathfrak{L}t'}[\rho(0)]j_x^P\}A_x(t')dt',
\end{align}
and the expectation value of the paramagnetic current is,
\begin{align}
\<j^P_x\>_t =& \text{Tr}[j_x^P\delta\rho(t)]\nonumber\\
=& i\int_{-\infty}^{t}\text{Tr}\{e^{\mathfrak{L}^\dagger(t-t')}[j_x^P] j_x^P e^{\mathfrak{L}t'}[\rho(0)]\nonumber\\
&-e^{\mathfrak{L}^\dagger(t-t')}[j_x^P]  e^{\mathfrak{L}t'}[\rho(0)]j_x^P\}A_x(t')dt'\nonumber\\
=& i\int_{-\infty}^t\text{Tr}
\{ [e^{\mathfrak{L}^\dagger(t-t')}[j_x^P],j_x^P]e^{\mathfrak{L}t'}[\rho(0)] \} A_x(t')dt'\nonumber\\
=& i\int_{-\infty}^{t}\< e^{\mathfrak{L}^\dagger t'}\{[e^{\mathfrak{L}^\dagger(t-t')}[j_x^P],j_x^P]\}\>A_x(t')dt'
\label{Eq:linearresponseformalismDissipation}
\end{align}
Eq.~\ref{Eq:linearresponseformalismDissipation} is a generalization of the conventional linear response formalism to the case with dissipation.

For the non-relativistic boson model, according to Eq.~\ref{Eq:atbdaggertDissipation}, \ref{Eq:bosonbilinearLindblad}, and \ref{Eq:nonrelativisticCurrent},

\begin{align} 
&[e^{\mathfrak{L}^\dagger(t-t')}[j_x^P],j_x^P]\nonumber\\
=&\sum_k\frac{4{e^*}^2}{m^2}k_x^2e^{-\Gamma(t-t')}(a_kb_{-k}u_k(t-t')v_k(t-t') -h.c.),
\end{align}
and for a probe field of frequency $\omega$

\begin{align}
\<j_x^P\>_t =& -\sum_k\frac{8{e^*}^2}{m^2}k_x^2A_x \int_{0}^{t}dt'e^{-i\omega t'} e^{-\Gamma(t-t')}\nonumber\\
&\text{Im}\{\<e^{\mathfrak{L}^\dagger t'}[a_kb_{-k}]\>u_k(t-t')v_k(t-t')\}\nonumber\\
=&-\sum_k\frac{8{e^*}^2}{m^2}k_x^2A_x \int_{0}^{t}dt'e^{-i\omega t'} e^{-\Gamma(t-t')}\nonumber\\
&\text{Im}\{[u_k(t')v^*_k(t')+m^\Gamma_k(t')]u_k(t-t')v_k(t-t')\}
\label{Eq:paramagneticcurrentDissipationFinal}
\end{align}
Note that $u_k(t')$ and $v_k(t')$ are sum of exponential functions of $t'$. By Eq.~\ref{mkGamma}, $m_k^\Gamma(t')$ is also a sum of exponential functions of $t'$. Furthermore, when $\omega\neq 0$, the integral in Eq.~\ref{Eq:paramagneticcurrentDissipationFinal} is a sum of exponential functions of $t$ whose growth exponent is no larger than $(2\theta_k - \Gamma)$. Thus the integral is bounded by $c_ke^{(2\theta_k - \Gamma)t}$, where $c_k$ is independent of time. For $\tilde{\lambda} t\gg 1, \omega t\gg 1$, the momentum summation is dominated by those momenta that $E_k\simeq\Omega/2$, and we can approximate the leading exponential function as 
\be 
e^{(2\theta_k-\Gamma) t}\simeq e^{(2\tilde{\lambda}-\Gamma) t} e^{ - \frac{(\Omega/2-E_k)^2}{\tilde{\lambda}^2}\tilde{\lambda} t}
\ee 
and approximate the momentum summation by a Gaussian integral of $(\Omega/2-E_k)/\tilde{\lambda}$. However
\be 
\text{Im}\{[u_k(t')v^*_k(t')+m^\Gamma_k(t')]u_k(t-t')v_k(t-t')\} = 0\nonumber
\ee 
when $E_k=\Omega/2$. Thus the leading contribution to Eq.~\ref{Eq:paramagneticcurrentDissipationFinal} is smaller compared to the diamagnetic current (see Eq.~\ref{Eq:diamagneticcurrentnonrelativistic}) by a factor of order $1/\tilde{\lambda} t$ or $1/\omega t$. Thus, in the limit $1/t\ll \omega,\tilde{\lambda}$, we conclude that $j_x\simeq j^D_x$ even with dissipation.

\section{Linear response of the relativistic boson}
\label{appendix:relativisticresponse}

In this appendix, we discuss the electromagnetic response of the relativistic boson model. We split the discussion into four sections. For simplicity , we take the boson velocity $v$ to be unity throughout this appendix unless otherwise specified.

\subsection{Lattice regularization}

For the relativistic boson model, the current operator is 
\be 
\vec{j}_\text{rel}(q)=\vec{j}^P_\text{rel}(q) + \vec{j}^D_\text{rel}(q),
\ee 

\begin{align}
\vec{j}^P_\text{rel}(q)&=\sum_ke^* (\vec{k}+\frac{\vec{q}}{2})\Psi_{-k}\Psi_{-k-q}^\dagger\nonumber\\
&=\sum_ke^*\frac{\vec{k}+\frac{\vec{q}}{2}}{\sqrt{E_kE_{k+q}}}(a_{-k}^\dagger + b_k)(a_{-k-q}+b_{k+q}^\dagger),
\end{align}

\begin{align}
\vec{j}^D_\text{rel}(q)&=-\sum_k{e^*}^2|\Psi_{-k}|^2\vec{A}(q)\nonumber\\
&=-\sum_k\frac{{e^*}^2}{E_k}(a_{-k}^\dagger+b_k)(a_{-k}+b_k^\dagger)\vec{A}(q)
\end{align}

The relativistic boson model is more difficult to handle than the non-relativistic model. Even without periodic drive, the relativistic boson model has an apparently divergent conductivity if we naively introduce a momentum cutoff. For example, for a uniform vector potential along the x direction,

\begin{align}
\<j^D_x\> = -\sum_k\frac{{e^*}^2}{E_k}\<b_kb_k^\dagger\>A_x
= -{e^*}^2\int_0^\Lambda\frac{d^2k}{(2\pi)^2}\frac{1}{E_k}A_x.
\end{align}
On the other hand
\begin{align}
\<j^P_x\> &= \sum_n|\<n|j_x^P|0\>|^2\frac{2}{E_n-E_0}A_x={e^*}^2\int_{0}^{\Lambda}\frac{d^2k}{(2\pi)^2}\frac{k_x^2}{E_k^3}A_x.
\end{align}
Naively, $j_x=j^D_x+j^P_x$ grows linearly with the momentum cutoff $\Lambda$. However, physically, a uniform vector potential is pure gauge, which should not induce any current. 

There are multiple ways to regularize this response. Since we are interested in a condensed matter system, the most natural way is to use a lattice regularization. We consider the following general Lagrangian

\be 
\mathcal{L} = \sum_k|\partial_t\Psi_k|^2 -F_k|\Psi_k|^2
\ee 
instead of $F_k=k^2+\Delta^2$, we demand $F_k$ to be periodic in the B.Z. and approaches $k^2+\Delta^2$ only for small $k$. Minimally couple this model with electromagnetic field, we find

\begin{align}
K_{xx}&\equiv -j_x/A_x=-j^D_x/A_x-j^P_x/A_x\nonumber\\
&=\int\frac{d^2k}{(2\pi)^2} \frac{-2}{2\sqrt{F_k}}(\frac{e^*}{2\sqrt{F_k}}\partial_{k_x}F_k)^2 +\! \int \frac{d^2k}{(2\pi)^2}\frac{{e^*}^2}{2\sqrt{F_k}}\partial_{k_x}^2F_k\nonumber\\
&= \int \frac{d^2k}{(2\pi)^2}\frac{{e^*}^2}{2}\partial_{k_x}(F_k^{-1/2}\partial_{k_x}F_k) \nonumber\\
&=0
\end{align}

For the numerical results presented in Fig.~\ref{Fig:latetimeRelativistic}, we use $F_k = \Delta^2+\frac{W^2}{4}(2-\cos(k_x)-\cos(k_y))$ with $\Delta=1,W=8$, which simulates a relativistic boson with velocity $v=2\sqrt{2}$ at small momenta.

For the analytic calculation of the driven boson model, because the leading order contribution comes only from the resonant region $E_k\sim\Omega/2$, we do not need to explicitly use any regularization scheme. We can simply ignore all terms that do not grow exponentially.

\subsection{Electromagnetic response for $\omega\gg \lambda/\Omega$}
\label{Appendix:relativisticResponsewlarge}

Even though the final result does not depend on whether $\omega$ is larger or smaller than $\lambda/\Omega$, The calculation of conductivity is particularly simple in the limit that the probe frequency is larger, $\omega\gg \lambda/\Omega$. We first look at the conductivity in this limit.

In the Heisenberg picture, under the rotating wave approximation (Eq.~\ref{Eq:canonicalQuantization}, \ref{Eq:atbdaggert}, \ref{Eq:ukvk}), 
\be 
\Psi_k(t) \simeq\frac{1}{\sqrt{E_k}}(W_k(t)a_k(0) + W^*_k(t)b_{-k}^\dagger(0)),
\ee 
where $W_k(t)\equiv e^{-i\Omega t/2}u_k(t) + e^{i\Omega t/2}v_k(t)$. Thus,

\begin{align}
\<j_x^D (t)\> &=-\sum_k\frac{{e^*}^2}{E_k}|W_k(t)|^2A_xe^{-i\omega t}
\end{align}
On the other hand, by the standard Kubo formula,

\begin{align}
\<j^P_x(t)\>=&\sum_k\frac{{e^*}^2k_x^2}{E_k^2}A_x\int_{-\infty}^{t}dt' e^{-i\omega t'}\nonumber\\
&i(W_k^2(t){W_k^*}^2(t')-W_k^2(t'){W_k^*}^2(t))
\label{Eq:parametricCurrentWkt}
\end{align}
We are interested in the limit $\omega,\lambda/\Omega\ll \Omega$. In the resonant region, we identify $u_k(t)$, $v_k(t)$ and $e^{-i\omega t}$ as slowly varying functions of time compared to $e^{\pm i\Omega t/2}$. Thus, we divide  $W^2_k(t)$ into three components

\begin{align}
W^2_k(t) =& e^{-i\Omega t}u^2_k(t) + e^{i\Omega t}v^2_k(t) + u_k(t)v_k(t)\nonumber\\
\equiv &W^2_{\Omega,k}(t) + W_{-\Omega,k}^2(t) + W_{0,k}^2(t)
\end{align}

Correspondingly, the current also has a slowly varying component and fast oscillating components. We rewrite the integral in Eq.~\ref{Eq:parametricCurrentWkt} as follows, keeping only the slowly varying component

\begin{align}
-\int_{-\infty}^{t}dt'e^{-i\omega t'}2\text{Im}[W_k^2(t){W_k^*}^2(t')] =C_k + D_k +\dots
\end{align}

\begin{align}
C_k\equiv & -\int_{-\infty}^{t}dt'e^{-i\omega t'} 2 \text{Im}[W_{\Omega,k}^2(t){W_{\Omega,k}^*}^2(t')]\nonumber\\
&-\int_{-\infty}^{t}dt'e^{-i\omega t'} 2 \text{Im}[W_{-\Omega,k}^2(t){W_{-\Omega,k}^*}^2(t')]\nonumber\\
D_k\equiv& -\int_{-\infty}^{t}dt'e^{-i\omega t'} 2 \text{Im}[W_{0,k}^2(t){W_{0,k}^*}^2(t')]
\label{Eq:CD}
\end{align}

In the limit $\theta,\omega\ll \Omega$, $C_k$ takes a simple form

\begin{align}
C_k\simeq  \frac{2}{\Omega}(|u_k(t)|^4 - |v_k(t)|^4)
= \frac{2}{\Omega}(2|v_k|^2+1),
\end{align}
where we have used $|u_k(t)|^2-|v_k(t)|^2 = 1$. Furthermore if we assume $\omega\gg\lambda/\Omega,\theta$, $D_k$ becomes negligible
\begin{align}
D_k = &iW_{0,k}^2(t)\int_{-\infty}^{t}{W_{0,k}^*}^2(t')e^{-i\omega t'}dt' \nonumber\\ &-i{W^*_{0,k}}^2(t)\int_{-\infty}^{t}W_{0,k}^2(t')e^{-i\omega t'}\nonumber\\
\simeq& -\frac{1}{\omega}|W_{0,k}^2|^2  +\frac{1}{\omega}|W_{0,k}^2|^2\nonumber\\
=&0
\label{Eq:B}
\end{align}

Combining the slowly varying component of the diamagnetic and the paramagnetic current, we find

\begin{align}
\<j_x(t)\>_\text{slow} &\simeq -\sum_k\frac{2{e^*}^2}{\Omega}(1-\frac{k_x^2}{E_k^2})(|v_k(t)|^2+1)A_x e^{-i\omega t}\nonumber\\
&\simeq\int\frac{kdk}{2\pi}\frac{{e^*}^2}{\Omega}(1-\frac{k^2}{2E_k^2})(\frac{\lambda}{2E_k\theta})^2e^{2\theta_k t}A_xe^{-i\omega t}\nonumber\\
&=\int\frac{EdE}{2\pi}\frac{{e^*}^2}{\Omega}(1-\frac{E^2-\Delta^2}{2E^2})(\frac{\lambda}{2E\theta})^2e^{2\theta t}A_xe^{-i\omega t}.
\end{align}

Define $x=\frac{\Omega/2-E}{\lambda/\Omega}$. $\theta \simeq\lambda/\Omega(1-x^2/2)$ and

\begin{align}
\<j_x(t)\>_\text{slow}&\simeq -\frac{\lambda}{\Omega}\frac{{e^*}^2}{8\pi}(1+\frac{4\Delta^2}{\Omega^2})e^{2\lambda t/\Omega}A_xe^{-i\omega t}\int dx e^{-x^2\lambda t/\Omega}\nonumber\\
&=-\frac{{e^*}^2}{8\sqrt{\pi}}(1+\frac{4\Delta^2}{\Omega^2})\sqrt{\frac{\lambda}{\Omega t}}e^{2\lambda t/\Omega}A_xe^{-i\omega t}
\end{align}

This result holds no matter whether $\omega\gg \lambda/\Omega$. However, if that condition is not satisfied, the contribution from $D_k$ is apparently nonzero, giving unphysical results, if we do not take into account corrections to the rotating wave approximation.

\subsection{Correction to the rotating wave approximation}

We reorganize the Heisenberg equation (Eq.~\ref{Eq:HeisenbergFullRelativistic}) into resonant and non-resonant terms

\begin{align}
\frac{d}{dt}\left (\begin{array}{c}
a_k(t)\\
\tilde{b}_{-k}^\dagger(t)
\end{array}\right)
=& -i\left (\begin{array}{cc}
E_k & \frac{\lambda}{2E_k}\\
-\frac{\lambda}{2E_k} & \Omega-E_k\\
\end{array}\right)
\left (\begin{array}{c}
a_k(t)\\
\tilde{b}_{-k}^\dagger(t)
\end{array}\right)\nonumber\\
&+i\frac{\lambda}{2E_k}\left (\begin{array}{cc}
-2\cos{\Omega t} & -e^{2i\Omega t}\\
e^{-2i\Omega t} & 2\cos{\Omega t}
\end{array}\right)
\left (\begin{array}{c}
a_k(t)\\
\tilde{b}_{-k}^\dagger(t)
\end{array}\right),
\end{align}
where $\tilde{b}^\dagger_k(t)\equiv e^{-i\Omega t}b^{\dagger}_k(t)$. In the resonant region $E_k\simeq \Omega/2$, $a_k$ and $\tilde{b}^\dagger_{-k}$ approximately oscillates near frequency $\Omega/2$. The first term on the right hand side oscillates at the same frequency as the left hand side; we call it the resonant term. The second term is the non-resonant term, which oscillates at frequencies differ to the bare frequency by $O(\Omega)$ if we substitute $a_k$ and $\tilde{b}_{-k}^\dagger$ by its bare solution at $\lambda = 0$. To the first order, we can ignore the second term

\begin{align}
\frac{d}{dt}\left (\begin{array}{c}
a_k(t)\\
\tilde{b}_{-k}^\dagger(t)
\end{array}\right)^{(1)}
=& -i\left (\begin{array}{cc}
E_k & \frac{\lambda}{2E_k}\\
-\frac{\lambda}{2E_k} & \Omega-E_k\\
\end{array}\right)
\left (\begin{array}{c}
a_k(t)\\
\tilde{b}_{-k}^\dagger(t)
\end{array}\right)^{(1)}
\end{align}

To the second order, the correction to the solution above satisfies

\begin{align}
\frac{d}{dt}\left (\begin{array}{c}
a_k(t)\\
\tilde{b}_{-k}^\dagger(t)
\end{array}\right)^{(2)}\!\!
=& -i\left (\begin{array}{cc}
E_k & \frac{\lambda}{2E_k}\\
-\frac{\lambda}{2E_k} & \Omega-E_k\\
\end{array}\right)
\left (\begin{array}{c}
a_k(t)\\
\tilde{b}_{-k}^\dagger(t)
\end{array}\right)^{(2)}\nonumber\\
&+i\frac{\lambda}{2E_k}\left (\begin{array}{cc}
-2\cos{\Omega t} & -e^{2i\Omega t}\\
e^{-2i\Omega t} & 2\cos{\Omega t}
\end{array}\right)
\left (\begin{array}{c}
a_k(t)\\
\tilde{b}_{-k}^\dagger(t)
\end{array}\right)^{(1)}
\end{align}
Thus,
\begin{align}
\frac{d}{dt}&\left (\begin{array}{c}
a_k(t)\\
\tilde{b}_{-k}^\dagger(t)
\end{array}\right)^{(2)}\!\!
=e^{-i\Omega t/2}\int_{0}^{t}dt'\frac{\lambda}{2E_k}\!\!\left(\!\begin{array}{cc}
  u_k(t-t')   & v^*_k(t-t')  \\
  v_k(t-t')   & u^*_k(t-t') 
\end{array}\!\!\right)\nonumber\\
&\left(\begin{array}{cc}
  -2i\cos{\Omega t'}  & -ie^{2i\Omega t'}  \\
   ie^{-2i\Omega t'}  & 2i\cos{\Omega t'}
\end{array}\right)
\left(\begin{array}{cc}
   u_k(t')  & v^*_k(t')  \\
   v_k(t') & u^*_k(t')
\end{array}\right)
\left (\begin{array}{c}
a_k(0)\\
\tilde{b}_{-k}^\dagger(0)
\end{array}\right)
\end{align}
Keep only terms at the order of $\lambda/\Omega^2$, we find

\begin{align}
\Psi_k^{(2)}(t) =&  W^{(2)}_k(t) a_k(0) + W^{*(2)}_k(t)b_{-k}^\dagger(0),\nonumber\\
W^{(2)}_k(t) =& \frac{\lambda}{\Omega^2}[e^{i\Omega t/2}(\frac{1}{2}u^*_k(t)-u_k(t)) + e^{-i\Omega t/2}(\frac12v^*_k(t)-v_k(t))\nonumber\\
&+\frac12e^{-3i\Omega t/2}u_k(t) + e^{3i\Omega t/2}v_k(t)] + O(\frac{\lambda^2}{\Omega^4}).
\end{align}

In the next section, we plug in the second order result
\begin{align}
W_k(t) = W^{(1)}_k(t) + W^{(2)}_k(t)
\end{align}
into Eq.~\ref{Eq:parametricCurrentWkt} to calculate the conductivity at arbitrary frequency.

\subsection{Electromagnetic response at arbitrary frequency}

In Eq.~(\ref{Eq:parametricCurrentWkt}-\ref{Eq:CD}), we write the parametric current as an integral involving $W^2_k(t)$ and identify contributions from $C_k$ and $D_k$. The contribution from $C_k$ involves the time integral of fast oscillating terms, which is suppressed by a factor of $\frac{1}{\Omega}$. The leading order $W_k(t)$ gives a contribution comparable to the diamagnetic current; the correction to $W_k(t)$ only gives a much smaller contribution. On the other hand, the contribution from $D_k$ is naively $\Omega^2/\lambda$ larger than the diamagnetic current. For consistency, we need to include corrections of $W_k(t)$ at the order of $\lambda/\Omega^2$.

Now we formally treat $\frac{\omega}{\lambda/\Omega}$ as an $O(1)$ number, and treat $\Omega/(\lambda t)$ and $\lambda/\Omega^2$ as small numbers. More specifically, note that $D_k$ depends on the momentum only through the energy and that $D_k$ is prominent only in the resonant region. Therefore, we define $x\equiv \frac{\Omega/2 - E}{\lambda/\Omega}$ and expand $D_k$ as a power series of $x$ near $x=0$. By explicit calculation, we find that

\begin{align}
D_k \simeq & (\frac{2\theta}{4\theta^2+\omega^2} - i\frac{4\theta^2}{(4\theta^2+\omega^2)\omega})e^{-i\omega t}\nonumber\\
&\cdot(x-2\frac{\lambda}{\Omega^2} + \frac{3}{2}x^3 -8\frac{\lambda}{\Omega^2}x^2)e^{2\theta t} \nonumber\\
&+\frac{i}{\omega}(x-2\frac{\lambda}{\Omega^2} + \frac{3}{2}x^3 -8\frac{\lambda}{\Omega^2}x^2)e^{2\theta t}\nonumber\\
&+\frac{2\theta-i\omega}{4\theta^2+\omega^2}(x-\frac{\lambda}{\Omega^2}+\frac{3}{2}x^3 -\frac{15\lambda^2}{2\Omega^2}x^2)e^{2\theta t}.
\label{Eq:Dkx}
\end{align}
We keep only terms which grow exponentially. The last two lines are non-oscillating in time; they give a DC component to the current. In the limit $\omega t\ll 1$, the probe field becomes pure gauge, and gauge invariance requires the DC component to cancel the AC component. However, the DC component is actually a response to the sudden turning on of the pump field at $t=0$. For probe fields with frequencies much larger than $1/t$, if we turn on the probe field after the pump (which is usually the case in experiments), the DC component disappears. Now we focus on the limit $\omega t\gg 1$. Note that

\begin{align}
    \theta \simeq \frac{\lambda}{\Omega}(1 -(x-x_0)^2/2),
\end{align}
where $x_0\equiv 2\lambda/\Omega^2$. For $\lambda t/\Omega\gg 1$, the momentum summation can be approximated as a Gaussian integral of x centered around $x_0$. After the Gaussian integral, $\<x\> \simeq x_0 = 2\lambda/\Omega^2$, $\<x^2\>\simeq \Omega/(2\lambda t)$, $\<x^3\>\simeq3/(\Omega t)$. The expansion of $D_k$ in $x$ becomes a mixed expansion in $\lambda/\Omega^2$ and $\Omega/(\lambda t)$. We ignore terms of $O((\lambda/\Omega^2)^2)$.

To the leading order in $\Omega/(\lambda t)$, we can safely ignore $x^2$ and $x^3$. Since $\<x\> = 2\lambda/\Omega^2$, we find from Eq.~\ref{Eq:Dkx} that the AC component is zero at this order. Thus we extend the validity of the results in Appendix~\ref{Appendix:relativisticResponsewlarge} to smaller frequencies.

To the next order in $\Omega/(\lambda t)$, we get a small correction to the superfluid density and a real part of the conductivity.

\begin{align}
    \text{Re}\ \sigma(\omega,t)=\frac{{e^*}^2}{16\sqrt{\pi}}(1+\frac{12\Delta^2}{\Omega^2})\frac{4(\frac{\lambda}{\Omega})^2}{(4(\frac{\lambda}{\Omega})^2 + \omega^2)\omega^2 t}\sqrt{\frac{\lambda}{\Omega t}}e^{2\lambda t/\Omega}.
\end{align}

Note that the real part is much smaller than the superconducting-like imaginary part when $1/t\ll \omega,\lambda/\Omega$. $\text{Re}\ \sigma(\omega, t)\sim O(1/(\omega t),\Omega/(\lambda t))\text{Im}\ \sigma(\omega, t)$.

\section{No Meissner effect at early time}
\label{appendix:Meissner}

In this appendix, we calculate the response of the non-relativistic boson model to magnetic field at early time, ignoring the dissipation, the boson interaction, and the decay of the periodic drive.

Consider a probe field
\be 
\vec{A} = A_x e^{iq_y y - i\omega t}\hat{x},
\ee 
which is smoothly turned on \textit{before} the pump. The paramagnetic current induced by this probe field is,

\begin{align}
\<j_x^P(q,t)\>=&\int_{-\infty}^{t}i\<[j_x^P(q,t),j_x^P(-q,t')]\>A_x e^{-i\omega t'}dt'\nonumber\\
=&-2\int_{-\infty}^{t}\text{Im}[\<j_x^P(q,t)j_x^P(-q,t')\>]A_xe^{-i\omega t'}dt'\nonumber\\
=&-\frac{4{e^*}^2}{m^2}\sum_k k_x^2A_xe^{-i\omega t}
\int_{-\infty}^{t}dt'e^{i\omega (t-t')}\nonumber\\
&\text{Im}[u_{k+q}^*(t')v_k^*(t')(u_{k+q}(t)v_k(t)+u_{k}(t)v_{k+q}(t))],
\label{Eq:paramagneticcurrentMeissner}
\end{align}
where $u_k(t)$ and $v_k(t)$ are given by Eq.~\ref{Eq:ukvknonrelativistic}.

We find that the integral on the last line of Eq.~\ref{Eq:paramagneticcurrentMeissner} grows at most by $e^{2\tilde{\lambda}t}$. In the limit $1/t\ll \omega, \sqrt{\Omega/m}q,\tilde{\lambda}$, to compute the leading-order behavior, we only need to keep terms proportional to $e^{2\theta_kt}$ and $e^{2\theta_{k+q}t}$, and approximate the momentum summation by an integral near $E_k = \Omega/2$ and $E_{k+q} = \Omega/2$ where the growth exponent reaches its maximum. Contributions from all other terms are suppressed by powers of $1/t$. For example, $e^{(\theta_k+\theta_{k+q})t}$ reaches its maximum only when $E_k=E_{k+q}=\Omega/2$; this much smaller phase space results in a much smaller contribution in the limit  $1/t\ll \sqrt{\Omega/m}q$. To further simplify the result, we utilize the freedom to interchange $k$ and $k+q$ in the momentum summation, and we take the limit $\omega\ll \tilde{\lambda}$. With these simplifications, we find that in 2D

\begin{align}
\<j_x^P(q,t)\> \simeq&\frac{{e^*}^2}{m^2}\sum_k \frac{k_x^2(E_{k+q}-E_k)}{(E_{k+q}-E_k)^2-2i\omega\tilde{\lambda}}e^{2\theta_kt}A_xe^{-i\omega t}\nonumber\\
=&\frac{e^*}{m}\int\frac{kdk}{2\pi}e^{2\theta_kt}\nonumber A_x e^{-i\omega t}\\
&\cdot\int \frac{d\phi}{2\pi}\frac{\sin^2\phi(qk_\Omega\cos\phi+\frac{q^2}{2})}{(q\cos\phi+\frac{q^2}{2k_\Omega})^2-2i\omega\tilde{\lambda}\frac{m^2}{k_\Omega^2}}\nonumber\\
=&-\<j_x^D(t)\>\int\frac{d\phi}{2\pi}\frac{\sin^2\phi(\frac{1}{\epsilon}\cos\phi +1)}{(\cos\phi +\epsilon)^2-ia},
\end{align}
where $\phi$ is the angle between $\vec{k}$ and $\vec{q}$ (which we choose to be in the y axis), $k_\Omega\equiv\sqrt{2\Omega m}$, $\epsilon \equiv q/2k_\Omega$, and $a\equiv\frac{\tilde{\lambda}}{\Omega}\frac{\omega}{q^2/2m}$. 

We find that the integral is singular in the small $q$ and $\omega$ limit; it approaches 0 for $a\gg 1$ and approaches $1$ for $a\ll 1$. Thus, for the probe field with frequency and momentum $\omega,q$ satisfying $\omega\ll \frac{\Omega}{\tilde{\lambda}}\frac{q^2}{2m}$, for example for a static magnetic field, there is no superconducting-like response at early time.
 
\end{document}